# The thermal, mechanical, structural, and dielectric properties of cometary nuclei after Rosetta

O. Groussin[1], N. Attree[1,2], Y. Brouet[3], V. Ciarletti[4], B. Davidsson[5],
G. Filacchione[6], H.-H. Fischer[7], B. Gundlach[8], M. Knapmeyer[9],
J. Knollenberg[9], R. Kokotanekova[10,11,12], E. Kührt[9], C. Leyrat[13],
D. Marshall[10,14], I. Pelivan[9,15], Y. Skorov[8,10], C. Snodgrass[11,16],
T. Spohn[9], F. Tosi[6]

[1] Aix Marseille Univ, CNRS, CNES, LAM, Laboratoire d'Astrophysique de Marseille, Marseille, France

[2] Earth and Planetary Observation Centre, University of Stirling, Stirling, UK

[3] Physics Institute, University of Bern, Bern, Switzerland

[4] LATMOS/IPSL, UVSQ (Université Paris-Saclay), UPMC (Sorbonne Univ.), CNRS, 78280 Guyancourt, France

[5] Jet Propulsion Laboratory, M/S 183-401, 4800 Oak Grove Drive, Pasadena, CA, 91109, USA

[6] INAF-IAPS, Istituto di Astrofisica e Planetologia Spaziali, Rome, Italy

[7] DLR Microgravity User Support Centre, Linder Höhe, 51147 Cologne, Germany

[8] Institut für Geophysik und extraterrestrische Physik, Technische Universität Braunschweig

[9] DLR Institute of Planetary Research, Rutherfordstr. 2, 12489 Berlin, Germany

[10] Max Planck Institute for Solar System Research, Justus-von-Liebig-Weg 3, 37077, Göttingen, Germany

[11] Planetary and Space Sciences, School of Physical Sciences, The Open University, Milton Keynes, MK7 6AA, UK

[12] European Southern Observatory, Karl-Schwarzschild-Strasse 2, 85748 Garching bei München, Germany

[13] Laboratoire d'Etudes Spatiales et d'Instrumentation en Astrophysique (LESIA), Observatoire de Paris, CNRS, UPMC, Paris Diderot Observatoire de Paris, 5 place Jules Janssen, 92195 Meudon, France

[14] Universität Göttingen, Friedrich-Hund-Platz 1, 37077 Göttingen, Germany

[15] Helmholtz Centre Potsdam, GFZ, German Research Center For Geosciences, 14473 Potsdam, Germany

[16] Institute for Astronomy, University of Edinburgh, Royal Observatory, Edinburgh EH9 3HJ, UK

**Corresponding author**

Olivier Groussin – Laboratoire d'Astrophysique de Marseille, 38 rue Frédéric Joliot-Curie, 13013 Marseille, France – Tel: +33 491 056 972 – Email: olivier.groussin@lam.fr

**Keywords**

*Solar System, Comets, Rosetta, Space Mission*

# Abstract

The physical properties of cometary nuclei observed today relate to their complex history and help to constrain their formation and evolution. In this article, we review some of the main physical properties of cometary nuclei and focus in particular on the thermal, mechanical, structural and dielectric properties, emphasizing the progress made during the Rosetta mission. Comets have a low density of 480 ± 220 kg m$^{-3}$ and a low permittivity of 1.9 – 2.0, consistent with a high porosity of 70 – 80 %, are weak with a very low global tensile strength <100 Pa, and have a low bulk thermal inertia of 0 – 60 J K$^{-1}$ m$^{-2}$ s$^{-1/2}$ that allowed them to preserve highly volatiles species (e.g. CO, $CO_2$, $CH_4$, $N_2$) into their interior since their formation. As revealed by 67P/Churyumov-Gerasimenko, the above physical properties vary across the nucleus, spatially at its surface but also with depth. The broad picture is that the bulk of the nucleus consists of a weakly bonded, rather homogeneous material that preserved primordial properties under a thin shell of processed material, and possibly covered by a granular material; this cover might in places reach a thickness of several meters. The properties of the top layer (the first meter) are not representative of that of the bulk nucleus. More globally, strong nucleus heterogeneities at a scale of a few meters are ruled out on 67P's small lobe.

# 1. Introduction

Comets were formed during the early stages of the solar system, and have been stored in the Kuiper belt and Oort cloud for billions of years, before they got injected into the inner solar system and ultimately became visible. Many processes have altered their nucleus since their formation, including: collisions, radiogenic and solar heating, encounters with giant planets, space weathering, phase transitions and erosion. A key question is to understand how the physical properties of cometary nuclei observed today relate to their complex history and more generally what it tells us about the formation and evolution of our solar system.

The physical properties of cometary nuclei are numerous, from basic attributes such as size, shape, albedo, or rotation period, to more complex properties such as the nature of the cometary material and its strength, porosity, or conductivity. Measuring these properties with a ground- or spaced-based telescope is challenging due to the small size of the nucleus, which is spatially unresolved, and its surrounding coma, which masks it. It has however been possible to derive the basic properties of more than a hundred cometary nuclei, from which the picture emerges of a nucleus with a kilometric size, a non-spherical shape, a low albedo (e.g. Lamy et al. 2004), a low density (e.g. Weissman et al. 2004) and a low conductivity (e.g. Fernández et al. 2013). To derive more accurate parameters, including surface heterogeneities, a dedicated space mission with a flyby or better a rendezvous offers the best solution. However, as shown in Fig. 1, this has only been possible for 6 cometary nuclei so far: 1P/Halley, 19P/Borrelly, 81P/Wild 2, 9P/Tempel 1, 103P/Hartley 2, and 67P/Churyumov-Gerasimenko (not mentioning the "blind" flybys of comets 21P/Giacobini-Zinner in 1985 and 26P/Grigg-Skjellerup in 1992).

In this paper, we review some of the main physical properties of cometary nuclei and focus in particular on the thermal properties (Sect. 2), the mechanical properties (Sect. 3), and the structural and dielectric properties (Sect. 4), emphasizing the progress made during the Rosetta mission of the European Space Agency.

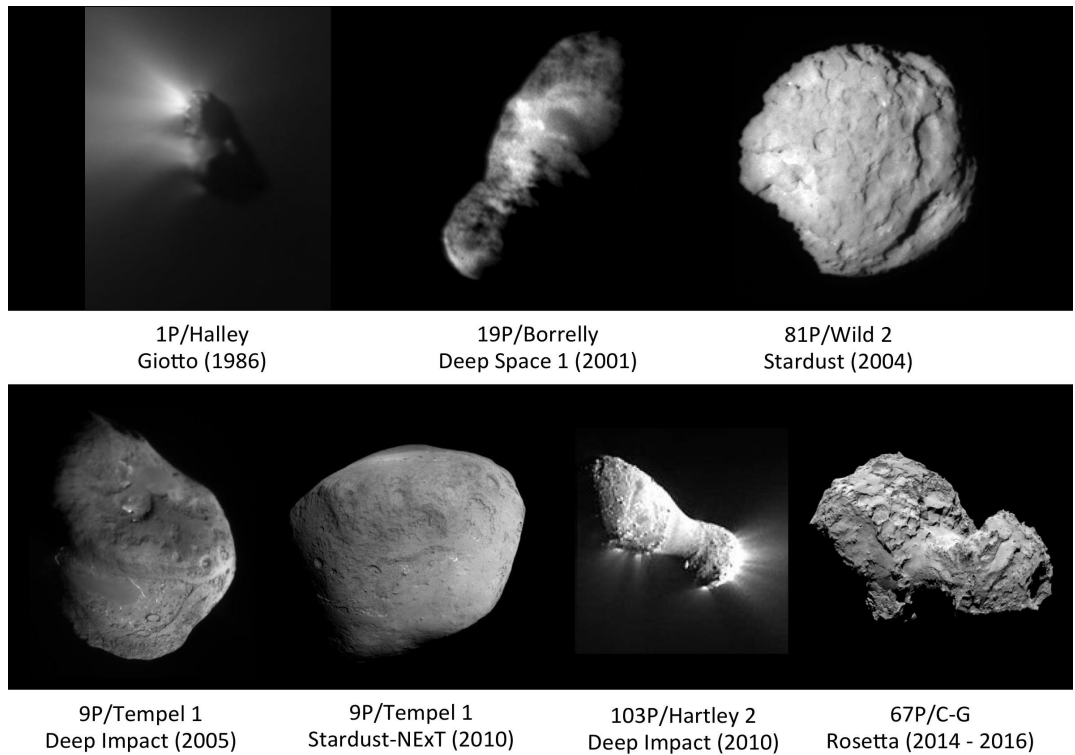

**Fig. 1** Images of the 6 cometary nuclei visited by spacecraft: 1P/Halley (credits ESA-MPAE), 19P/Borrelly (credits NASA/JPL), 81P/ Wild 2 (credits NASA/JPL), 9P/Tempel 1 (credits NASA/JPL/UMD and NASA/JPL/Caltech-Cornell), 103P/Hartley 2 (credits NASA/JPL/UMD), and 67P/Churyumov-Gerasimenko (credits ESA/Rosetta/MPS for OSIRIS Team MPS/UPD/LAM/IAA/SSO/INTA/UPM/DASP/IDA).

## 2. The nucleus thermal properties

The thermal properties drive the ability of the comet nucleus to response to solar illumination and define how heat is transported inside the nucleus material. They define the temperature of the nucleus and are therefore key to properly describe many physical processes occurring on it, including the surface thermal emission, the sublimation of ices and their recondensation, the conduction of heat into the nucleus porous media, phase transitions and the transport of gas. The thermal properties are derived from temperature measurements (Sect. 2.1), and best described by the thermal inertia (Sect. 2.2).

## 2.1. Temperature

### 2.1.1. State of the art before Rosetta

Before the spacecraft flybys at comet 1P in 1986, the nucleus surface temperature was mainly constrained by modelling, assuming a mixture of ice and dust, possibly covered by a dust mantle of varying thickness (e.g. Whipple 1950, Mendis and Brin 1977, Fanale and Salvail 1984). At 1 au from the Sun, the typical surface temperature at the sub-solar point was therefore between ~200 K, the free sublimation temperature of water ice, and ~400 K, the black body temperature, without being able to favour one case to the other.

Following the 1P flyby, the picture emerged of a monolithic and dark nucleus, with active and inactive regions, each of them with a different temperature (Julian et al. 2000). The temperature measured at the surface of 1P by the IKS instrument onboard the VEGA mission was larger than 360 K at 0.8 au from the Sun (Fig. 2), suggesting that the surface is mainly covered by hot inactive materials, depleted in volatiles, and with a low thermal conductivity because of the temperature close to the black body temperature (Emerich et al. 1987).

Progress occurred in the 2000's with the in-situ temperature measurements of four cometary nuclei, moreover spatially resolved (Fig. 2). Results indicate that the nucleus reaches its maximum temperature close to the subsolar point, and that the maximum temperature is close to the black body temperature; these two effects show that the thermal inertia of the nucleus is small enough not to cool the surface temperature significantly or to shift it temporally relatively to the subsolar point (Sect. 2.2). For comets 9P and 103P, the temperature spatial distribution shows no correlation with exposed water ice on the surface, which is therefore not present in sufficient amount to cool it (Sunshine et al. 2006). Finally, the colour temperature decreases by only a few tens of Kelvin from the subsolar point to solar incident angles exceeding 60°, which demonstrates that the surface is rough at the instrument sub-pixel scale (<10 m) (Sect. 4.1; Groussin et al. 2013).

Concerning the temperature inside the nucleus, it is only constrained by modelling. Assuming that the comet nucleus is characterised by a low thermal conductivity (Sect. 2.2) and high porosity (Sect. 4.3), temperature decreases rapidly with depth on the day side and only a small fraction of the surface solar energy is transported into the nucleus at depths exceeding 1 m. As a consequence, simple thermodynamic estimates show that the Jupiter Family Comets (JFCs) observed today are out of thermal equilibrium, with a thermal relaxation time (~$10^5$ yr; see Prialnik 2004, with the parameters of 67P) almost one order of magnitude larger than their mean lifetime in the inner solar system (~$1.2 \times 10^4$ yr; Levison et al. 1997). On the contrary, their expected lifetime of ~$4 \times 10^8$ yr in the primordial disk around 15 – 30 au (Morbidelli et al. 2012; Marchi et al. 2013) and of ~$4.5 \times 10^7$ yr in the Kuiper belt around 40 au (Levison et al. 1997), is more than two orders of magnitude larger than the thermal relaxation time, which means that they should be at thermal equilibrium at these distances. It is therefore expected that the core temperature might be low, explaining the presence of volatile species (e.g. $H_2O$, $CO_2$) in comets, but possibly above the sublimation temperature of super volatiles (e.g. CO, $N_2$), which could therefore be trapped in amorphous ice or clathrates (Mousis et al. 2015).

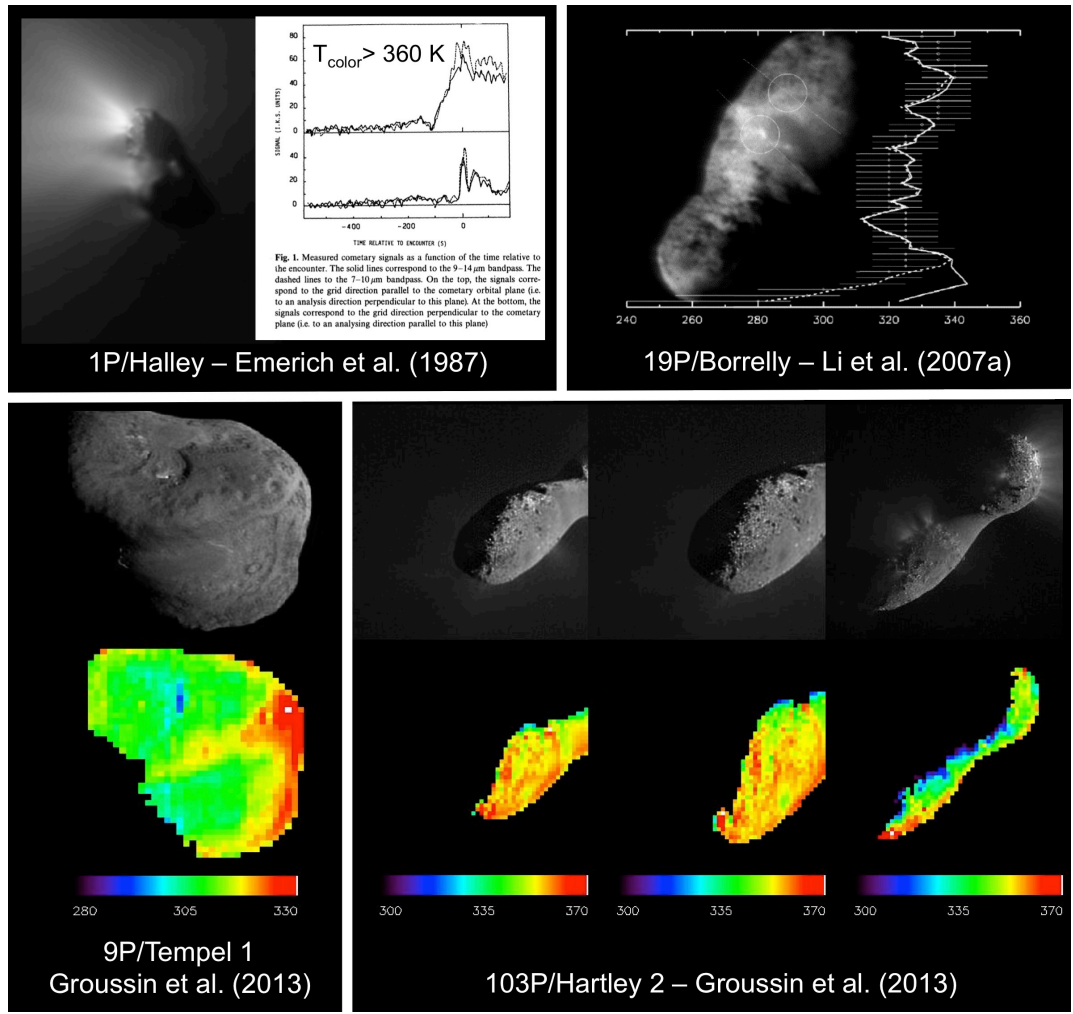

**Fig. 2** Spacecrafts color temperature measurements of four cometary nuclei, before the Rosetta mission.

**2.1.2. Progress during Rosetta**

*2.1.2.1. VIRTIS temperature measurements*

Temperature maps of the nucleus of comet 67P have been derived by modelling its thermal emission on spectral radiance data measured by VIRTIS-M (Visible InfraRed and Thermal Imaging Spectrometer, Mapping channel), the VIS-IR hyperspectral imaging channel aboard Rosetta (Coradini et al. 2007). The instrument performs 0.25 – 5.1 µm imaging spectroscopy on two separate channels, covering the 0.25 – 1.0 µm and 1.0 – 5.1 µm ranges respectively, by means of two bidimensional detectors, a CCD and an MCT array, sharing the same telescope equipped with a scan mirror to build hyperspectral images (so-called "cubes"). The average spectral sampling is equal to 1.8 nm and 9.7 nm per band for the two channels. On 3 May 2015, a sudden malfunction of the VIRTIS-M infrared focal plane cryocooler inhibited the use of infrared data acquired by the mapping channel, for which only the visible channel (0.25 – 1 µm) was still operating. The infrared spectral range 2 – 5 μm was covered by the high-spectral resolution channel VIRTIS-H, with however no imaging capability, for the rest of the mission. VIRTIS-H is mainly devoted to observe the coma.

VIRTIS could sample temperatures within the uppermost radiatively active surface layer, typically tens of microns due to the spectral range being limited to ≤5.1 μm, the very low surface albedo and the high thermal emissivity. Within this wavelength range, the gray-body radiance is a non-linear function of temperature. Moreover, due to the typical VIRTIS pixel resolution values achieved on the nucleus (2 – 25 m $px^{-1}$), within a given resolution cell, the measured radiance is preferentially weighted by the warmest sub-pixel areas. This effect is relevant for the case of 67P due to the rough morphology of the surface (Sect. 4.1).

Surface temperatures, as retrieved by VIRTIS-M and covering the early pre-perihelion period from 1 August to 23 September 2014, are widely reported by Tosi et al. (2019). In this period, 67P was rapidly approaching the frost line, with the heliocentric distance decreasing from 3.62 to 3.31 au, and the spacecraft was in the altitude range 61 – 13 km above the surface, resulting in a spatial resolution from ~15 to ~3 m $px^{-1}$ (most data showing a resolution of 13 to 15 m $px^{-1}$). The solar phase angle ranged from 17° to 93°, which also allowed variable illumination and local time conditions to be explored. Due to the large obliquity of the comet (52°, Jorda et al. 2016) that causes pronounced seasonal effects, at the time of these observations Northern latitudes were in the summer season, while Southern latitudes below 45°S were in the winter season and experienced polar night (Tosi et al. 2019).

The lower limit of temperatures sensed by VIRTIS-M is controlled by the instrumental noise-equivalent spectral radiance (NESR), which may vary over time depending on some instrumental parameters. In the above mentioned time period, this lower bound was 156 K on average, which in fact restricts the VIRTIS-M coverage to the dayside of the nucleus, including areas that recently underwent shadowing or recently exited from shade (Tosi et al. 2019).

When filtering VIRTIS-derived thermal maps into different local time intervals, each covering two hours in true local solar time, the surface of comet 67P does not exhibit outstanding thermal anomalies. These maps show a fairly uniform distribution of surface temperatures, mildly dependent on the latitude. The big lobe and the "neck", i.e. the surface area connecting the two lobes display regional differences despite their mutual proximity, with the neck exhibiting a faster temperature change than the big lobe, both in the morning and in the evening hours. Finally, the regions in the big lobe that experienced grazing sunlight around local noon during this season show the lowest values of maximum daytime temperature. Figure 3 shows one example of such a map obtained for the 10 – 12 h true local solar time interval (Tosi et al. 2019).

Different morphological regions (Fig. 4; Thomas et al. 2018) show peculiar thermal behaviour: Ash, Babi, Hapi, and Seth regions, located near the neck, and the Ma'at region on the small lobe close to the neck, are the locations where maximum surface temperatures were detected, whereas other regions at comparable latitudes reached lower temperatures. This behaviour is correlated to favourable illumination conditions occurring at the time of VIRTIS observations. Besides increasing diurnal surface temperature, solar illumination drives the activity occurring in some of those areas (Keller et al. 2017), and particularly in the Hapi region (De Sanctis et al. 2015).

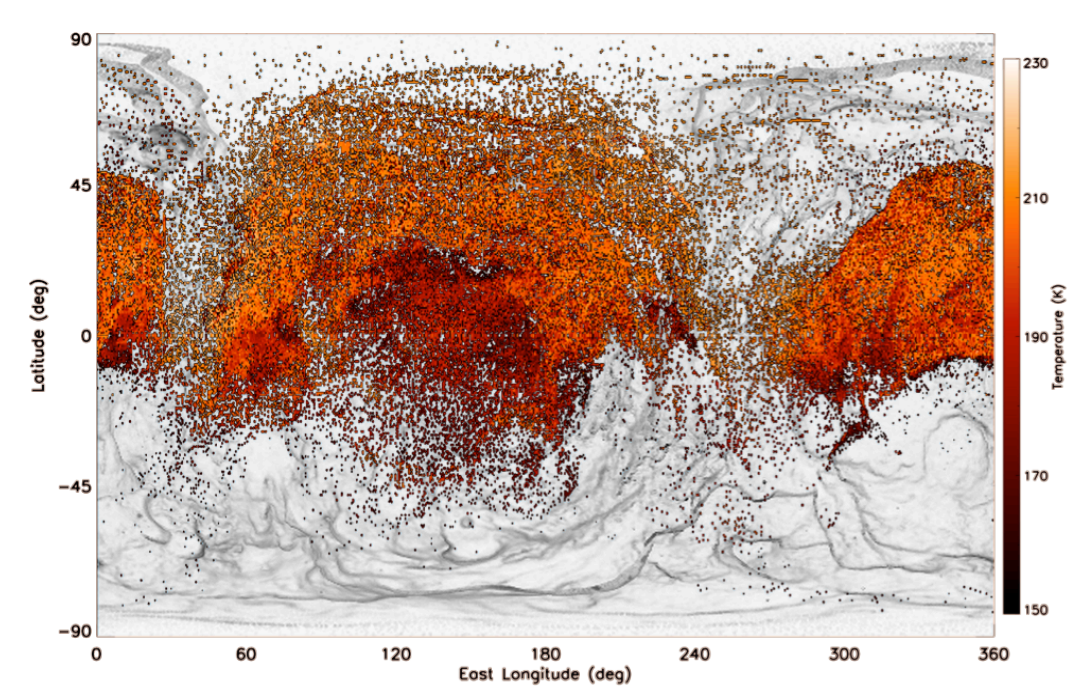

**Fig. 3** Temperature map of the nucleus of comet 67P obtained by VIRTIS between 10 h and 12 h true local solar time from 1 August to 23 September 2014, in cylindrical projection and with a fixed angular binning of 0.5° by 0.5° in both latitude and longitude. Each bin collects the average temperature recorded during a full diurnal rotation period. We display only illuminated points that have at the same time: solar incidence angle $i$<80°, emission angle $e$<80°, and accuracy better than 2 K. In this period, circumpolar regions experienced permanent lighting. For comparison, at that heliocentric distance, the black body temperature is around 215 K.

The temperature recorded on the dayside of the nucleus, for solar incidence angles <20° (i.e. around local noon) and emission angles <80°, is 213 ± 3 K on average (Tosi et al. 2019). No correlation arises on a global scale between VIRTIS-derived temperature and single scattering albedo, both at visible and near-infrared wavelengths (Ciarniello et al. 2015). Extreme values, reaching up to 230 K, were recorded by VIRTIS at Northern latitudes in the Seth region corresponding to the pit area named "Seth_01" (Vincent et al. 2015), and are most likely due to the peculiar topography of the pit combined with the instantaneous solar illumination shortly after local noon, which enhances the self-heating effect (Tosi et al. 2019). Self-heating results from the irregular shape and topography, which induces mutual heating of regions (or facets on a shape model) seeing each other.

When correlating the measured surface temperature values for all of the morphological regions in the sunlit hemisphere with the solar incidence angle, a strong dispersion is observed. This result indicates that most data points do not follow a black body behaviour. In the case of regions located in the neck area, such deviations can be attributed to macroscopic self-heating effects caused by the presence of large concavities. Such morphologies, along with the low surface albedo, causes local infrared thermal flux enhancement by repeated emissions from mutually facing surface areas. On the other hand, in the equatorial region the presence of several distinct trends in temperature vs. solar incidence angle could be ascribed to the non-uniformity of the morphology within a given region, and thus to a variable role of the mutually illuminating areas. Finally, temperature values higher than the ideal case of a black body, observed in all of the explored morphological regions for large solar incidence angles, are evidence for small-scale roughness (Sect. 4.1) (Tosi et al. 2019).

On 22 August 2014, VIRTIS obtained 7 consecutive snapshots of the nucleus, from an altitude of 60 km above the surface (spatial resolution 15 m px$^{-1}$). This sequence covers large areas of the neck and the two lobes, following them during ~15% of the rotational period, which allowed direct computation of thermal gradients throughout the scene. From these data, it emerges that the dayside of the nucleus shows typical temporal thermal gradients of 0.1 K min$^{-1}$, except in the neck area where sudden daytime shadowing takes place. In this latter case, the temporal gradient increases up to 2.0 K min$^{-1}$, i.e. twenty times larger than the typical value. However, the real temporal gradient in the neck is likely several times larger than this value, due to the poor ability of VIRTIS in retrieving surface temperatures <160 K (Tosi et al. 2019).

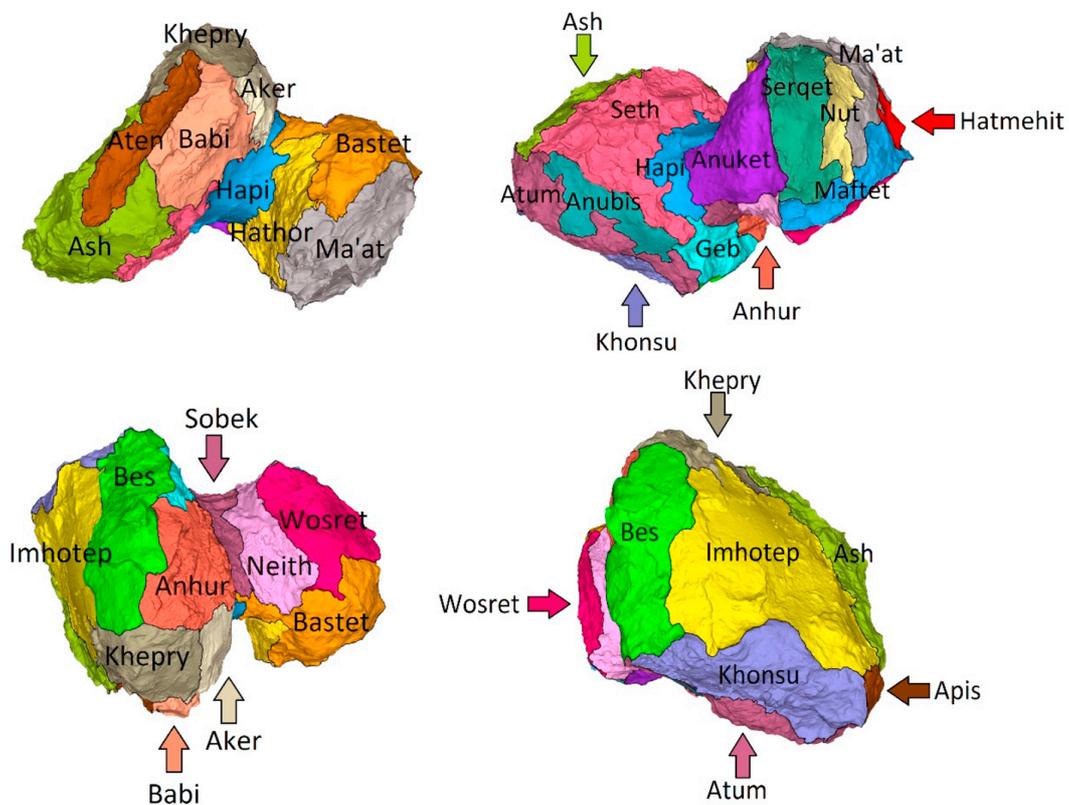

**Fig. 4** Regional units of the nucleus of 67P, from Thomas et al. (2018).

*2.1.2.2. MIRO temperature measurements*

The Microwave Instrument for the Rosetta Orbiter (MIRO) is a microwave spectrometer with two heterodyne receivers operating at 190 GHz (mm channel) and 560 GHz (sub-mm channel) (Gulkis et al. 2007). Each receiver measures an antenna temperature, which corresponds to the temperature from a sub-surface depth. The sub-mm channel typically probes the first cm, while the mm channel probes the first 4 cm (Schloerb et al. 2015). The measured antenna temperatures are converted to a brightness temperature, which is defined as the required temperature of a black body that fills the observed area to produce the observed power. In this section, whenever we refer to a temperature measured by MIRO, it is therefore a *brightness temperature*.

Measured temperatures from MIRO have been quoted in a variety of sources, although many of them only cover the very beginning of the mission, when 67P was beyond 3 au from the Sun. At the MIRO wavelengths, the measured temperatures originate from a subsurface (a few cm) volume element, not from the surface itself.

The earliest temperature measurements come from Gulkis et al. (2015), who reported results from 2014, showing diurnal and seasonal temperature variations. The data were restricted to a longitude band between 100° and 200° in order to avoid the neck region, where shadowing makes it difficult to interpret the diurnal phase curves. They show the temperature variation in latitude bands at 20°S – 30°S and 20°N – 30°N. During this early part of the mission, the Northern side was mainly illuminated. In the sub-mm channel, on the Southern side, the sub-surface temperature varied between approximately 60 – 100 K at midnight but between approximately 50 – 160 K at midday. In the well illuminated Northern hemisphere, the variation was far less dramatic, varying between 115 – 125 K at midnight and between 160-190 K at midday. In the mm channel, low sub-surface temperatures of 60 K were recorded at all times in the Southern hemisphere with highs of 110 K at midnight and 155 K at midday. In the Northern hemisphere, the sub-surface temperature varied between 140 – 155 K at midnight and 160 – 185 K at midday.

Temperatures from August 2014 are also reported by Lee et al. (2015) and show a similar behaviour: the sub-surface temperature on the well illuminated dayside reaches over 150 K, but the in less illuminated regions, it goes down to about 75 K.

Additionally, Gulkis et al. (2015) suggest that there is a slight temporal lag, with the highest sub-surface temperatures in the Southern hemisphere observed in the early or mid-afternoon, revealing the sensitivity to the diurnal heat cycle. Furthermore, the reduction in temperatures on the Southern side compared to the Northern side shows the seasonal effect on the comet.

The diurnal variation in sub-surface temperature in September 2014 is further investigated by Schloerb et al. (2015). In this month, the comet was at heliocentric distances of 3.45 – 3.27 au and the sub-solar point in mid-September was at 43°N. The observations are mainly of the Imhotep and Ash regions, and the measurements are binned by latitude and local solar time. Schloerb et al. (2015) find that the sub-surface temperature is a function of latitude, arising from the fact that solar flux is a function of latitude. The diurnal temperature curve in the sub-mm channel has a greater amplitude than the temperature curve of the mm channel. In both channels, the maximum is shifted after midday, with a larger shift for the mm channel compared to the sub-mm channel. The shift is greater for latitudes farther from the sub-solar point. These results indicate that the mm emission arises from greater depths than the sub-mm emission, and Schloerb et al. (2015) find that for the mm channel, the sub-surface temperature is measured over the first 4 cm of the nucleus surface, whereas for the sub-mm channel, the measured sub-surface temperature typically originates from the first centimetre.

A dedicated mapping scan of 67P was performed on 7 September 2014, allowing the temperature of the subsurface to be mapped across the whole nucleus. The scanning procedure took three hours and the results of the sub-mm mapping are shown in Biver et al. (2015). At this point in time, the variation in temperature across the comet is clearly evident, with the illuminated dayside having temperatures of approximately 160 K and as low as 20 K on the night side. In their work, Biver et al. (2015) also calculate the water column densities around the comet, finding the highest densities occur around the regions with the highest temperature, although the neck region is distinctly more productive than the hotter regions surrounding it (Marschall et al. 2018). This suggests that topography plays a key role in outgassing activity, in addition to solar insolation (Vincent et al. 2018).

Choukroun et al. (2015) used MIRO to assess the night-time temperatures on 67P from data taken between August 2014 and October 2014. The night-time sub-surface temperatures varied from 17 – 40 K in the sub-mm channel and 30 – 50 K in the mm channel. The mm channel probes deeper than the sub-mm channel, so the measured brightness temperature is a little higher than in the sub-mm channel, due to thermal inversion curve on the night side.

Beyond the early phase of the mission, continuum temperatures from MIRO covering August 2014 to April 2016 are given in Marshall et al. (2017). The observed sub-surface temperatures vary from 70 K to 255 K, with the highest temperatures recorded around perihelion when 67P was at 1.24 au from the Sun. In these measurements, it is unlikely that MIRO observed the hottest regions on the fully illuminated dayside as the spacecraft was in a terminator orbit during perihelion and so the maximum sub-surface temperature of the nucleus is probably higher than these recorded values.

*2.1.2.3. Philae temperature measurements*

Philae came to rest at the final landing site Abydos on 12 November 2014, 17:31:17 UTC (t=0 in the following text). Nine months before perihelion, the distance to the Sun was 3 au Several experiments and sensors on Philae provide information about the thermal environment at Abydos. These are the MUPUS (MUlti-PUrpose Sensors for surface and subsurface science) sensors MUPUS-TM (Thermal Mapper) and MUPUS-PEN (PENetrator) (Spohn et al. 2007), the SESAME (Surface Electric Sounding and Acoustic Monitoring Experiment) (Seidensticker et al. 2007) temperature sensors located in the feet of the lander, and Philae's housekeeping sensors and solar array data. Temperature data from the scientific sensors were not continuously acquired but only during certain operational phases. The most extended data set comes from MUPUS-TM that was operational during Block-1 (t = 0 – 11 h), the 4 safe blocks (t = 13 – 21 h), and the final operational Block-6 (t = 29 – 41 h), whereas the PEN was only operated in Block-6 (for a description of the operational blocks on the comet see Ulamec et al. 2016). SESAME temperature data were collected during the four safe blocks and in Block-6.

Figure 5 shows the diurnal cycle of the acquired temperatures after landing in Abydos. Data from different rotations are plotted as a function of local time. Here, we only show data that were not significantly affected by operational issues (i.e. due to self-heating) of the sensors and, therefore, provide real information about the thermal environment including the illumination history. This excludes SESAME data recorded with active accelerometers/transmitters and PEN data acquired before hammering was finished and the deployment boom retracted as well as all PEN sensors located in the upper 2/3 of the PEN. Because the PEN was not able to penetrate into the surface (Spohn et al. 2015), it was either sticking with its tip in the ground or lying somewhere on the ground close to the big boulder behind the Philae balcony. Therefore, the measured PEN temperatures are mainly reflecting an average of its radiative environment. The SESAME sensors provide good information about the illumination at Abydos but it is difficult to judge how well they measure the temperature of the soil beneath the soles. Knapmeyer at al. (2018) showed that all lander feet were in contact with the ground at Abydos but the reaction to illumination is certainly different between the feet and the cometary ground. Detailed modelling of this problem has not yet been done.

With respect to the MUPUS-TM temperatures it should be noted that the temperatures shown in Fig. 5 are somewhat higher than those given by Spohn et al (2015). The reason is a mistake in Eq. (2) of the supplementary material of Spohn et al (2015) that relates the measured thermopile signal to the kinetic temperature. A corrected version of this equation and a discussion of the underlying assumptions can be found in Grott et al. (2017). The consequences of this correction are that no estimate of the comet emissivity is possible from a single TM channel, and that the plotted brightness temperature is a lower bound for the surface temperature. The uncertainty introduced by the unknown emissivity is small for the low target temperatures at Abydos. For the broadband TM channel used here deviations are below 2.2 K if the emissivity is larger than 0.9. This uncertainty is incorporated in the error bars shown in Fig. 5.

The measured temperatures are consistent with surface temperatures between about 100 K and at least 165 K varying over time and with the exact position in Abydos. The temperatures of the feet show that first the +X-foot (towards the observer in the OSIRIS image of the lander at Abydos, Ulamec et al. (2017)) of the lander is illuminated, then the +Y foot, and that the -Y foot does not receive any direct illumination at all. The latter sensor gives nearly constant temperature readings of about 115 K, which should reflect the average temperature of the immediate environment of the -Y foot. It is further noteworthy that there is no discontinuity in the rising temperature curve of the SESAME sensors although different parts of the curve were recorded about 2 comet rotations apart, during Block-6  (Nov. 14, $t_{loc}$ = 2.7 – 3.7 h) and the first safe block (Nov. 13, $t_{loc}$ = 3.7 – 4.7 h). This is an indication that no significant movement of the lander occurred in between. The temperatures measured by TM and PEN start slightly above 100 K and then rise slowly over 4 h to reach about 120 K. This temperature rise is probably due to diffuse radiation (reflected visible and thermal infrared) reaching the TM target area and the PEN surroundings after sunrise in the Abydos region. This interpretation is supported also by lander solar array data. Here, it should be noted that due to the design of the Philae solar power system the output voltage of the solar cells is more sensitive to weak illumination levels than the generated current. The output voltage summed up over all solar panels rises at $t_{loc}$ ≈ 0 to approximately 10 V indicating that scattered visible light reaches the Philae hood.

At $t_{loc}$ ≈ 4 h the TM signal shows a steep increase to a maximum of 130 K whereas the PEN temperature continues to rise slowly. This was interpreted by Spohn et al. (2015) to indicate that a part of the TM FOV receives direct sunlight at this point in time whereas the PEN remains in shadow.  At $t_{loc}$ ≈ 4.7 h (dotted vertical line in Fig. 5) the night in Abydos begins as indicated by the solar array voltage dropping to zero. This is also reflected in the sharp decrease of the TM signal and in the temperature maximum reached at the +Y foot and the subsequent decay of all temperatures over the remaining part of the rotational period.

Independent of the diurnal cycle, there seems to exist a trend to lower temperatures with increasing time. TM temperatures recorded directly after landing (Block-1) are higher than the values measured one (safe block) and three (Block-6) rotational periods later, and also the temperature of the -Y foot recorded late during the first science sequence (in the diurnal curve between $t_{loc}$ = 2.7 h and 3.7 h) is lower than during earlier measurements (though this is not fully conclusive due to the large error bar). Such a trend to lower temperatures could be a consequence of additional shadowing of Abydos by the lander itself.

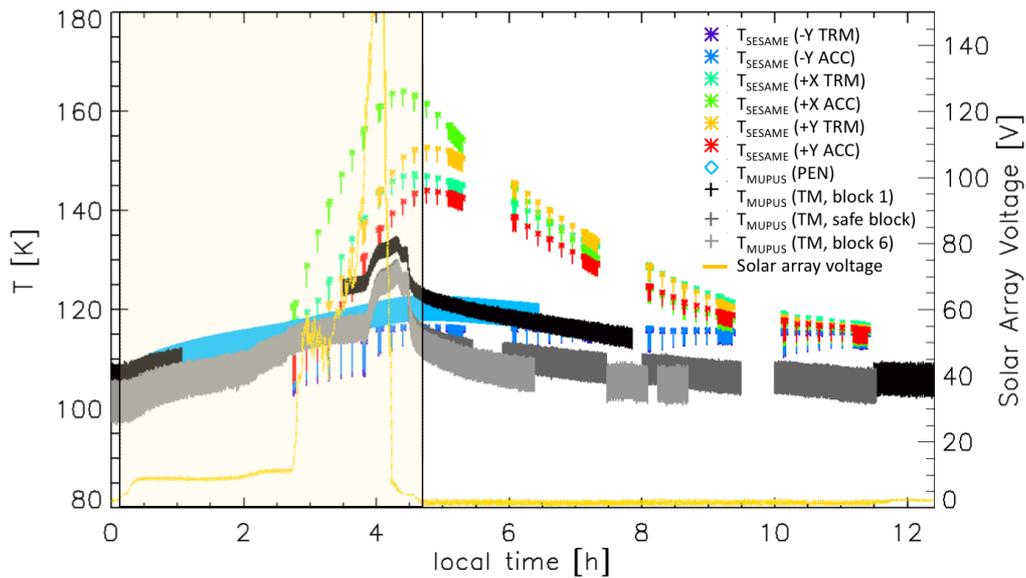

**Fig. 5** Temperatures of MUPUS-TM (brightness temperature of 6 – 25 µm broadband channel), MUPUS-PEN, and SESAME transmitters (TRM) and accelerometers (ACC) located in the three feet of Philae (designated as +X, -Y and +Y). Additionally, the solar array voltage (yellow line) is plotted as function of local time. The semi-transparent yellow box indicates the length of the local day (~4.7 hours).

## 2.2. Thermal inertia and conductivity

### 2.2.1. State of the art before Rosetta

The thermal inertia drives the ability of a material to adapt its temperature to a change in local insulation, either resulting from day/night variations or shadowing effects. A material with a large thermal inertia takes longer to adapt its temperature compared to a low thermal inertia material. The thermal inertia $I$ (in J K$^{-1}$ m$^{-2}$ s$^{-1/2}$) is defined as $I = \sqrt{\kappa \rho c}$, where $\kappa$ is the thermal conductivity (in W K$^{-1}$ m$^{-1}$), $\rho$ the density (in kg m$^{-3}$), and $c$ the heat capacity (in J K$^{-1}$). Since the thermal conductivity and the heat capacity are temperature dependent, the thermal inertia also depends on temperature; the thermal inertia of a given body changes depending on its heliocentric distance, which is particularly relevant for comets with highly elliptical orbits.

Estimating the thermal inertia has always been challenging, and until the flyby of comet 9P in 2005, mainly relied on in-situ (Vega) or remote infrared (ISO and Spitzer) and millimetre (IRAM) spatially unresolved observations of the nucleus to derive, at minimum its temperature, better the deviation of its spectral energy distribution from that of a black body, and even better a temporal shift in its thermal infrared light curve.

On comet 1P, the hot nucleus (Sect. 2.1.1), with a temperature close to that of a black body, suggested a low thermal inertia with negligible energy penetrating into the nucleus by heat conduction. From ISO observations at four wavelengths in the range 7 – 15 µm, Jorda et al. (2000) derived a negligible thermal inertia for comet C/1991 O1 Hale-Bopp, as well as Boissier et al. (2011) who derived a thermal inertia of 10 J K$^{-1}$ m$^{-2}$ s$^{-1/2}$ for comet 8P/Tuttle using Plateau de Bure observations. Fernandez et al. (2013) performed a survey of 57 cometary nuclei, using Spitzer two-wavelengths observations, and also concluded that the nucleus thermal inertia is negligible. More precisely, we converted their measured beaming parameter of 1.03 ± 0.11 to a thermal inertia of 30 ± 20 J K$^{-1}$ m$^{-2}$ s$^{-1/2}$, assuming a reasonable small-scale roughness ($\eta$ = 0.8; see Lebofsky et al. 1986 and references therein for a discussion of the $\eta$ parameter). For this purpose, we assumed a spherical nucleus at 4.5 au from the Sun (where most Spitzer data were acquired) with no thermal inertia and $\eta$ = 1.03, and computed its flux at 16 μm and 22 μm (Spitzer observational wavelengths); we then looked for the thermal inertia that best reproduces these fluxes, for the same nucleus but with $\eta$ = 0.8. Finally, from Spitzer thermal infrared light curves of 67P and 9P, Lamy et al. (2008) and Lisse et al. (2009) respectively, came to the same conclusion, with a thermal inertia <50 J K$^{-1}$ m$^{-2}$ s$^{-1/2}$.

A second, less direct, approach has also been used, which is to derive the thermal inertia from the porosity, the latter being derived from density estimates (Sect. 4.2). Since the nucleus density can be inferred from non-gravitational or tidal perturbations, this method is the oldest one, but the least accurate. For example, in the 80's and 90's, the density was estimated to 600 kg m$^{-3}$ for the nucleus of 1P (Sagdeev et al. 1988) and to 300 – 700 kg m$^{-3}$ for that of D/1993 F2 Shoemaker-Levy 9 (SL9) (Asphaug and Benz 1994). This low density suggested a low conductivity, therefore a low thermal inertia, but the analysis is model dependent, therefore more qualitative than quantitative.

In parallel to the above observations, an effort has been made to constrain the thermal inertia of porous icy materials from laboratory experiments and modelling. From the experimental point of view, we can mention the KOSI experiments (Huebner 1991; Kochan et al. 1998) on porous ice/dust samples with a measured thermal conductivity of 0.3 - 0.6 W m$^{-1}$ K$^{-1}$, ten times lower than that of compact ice (Spohn et al. 1989). More recently, Krause et al. (2011) built an experiment, combined with numerical simulations, to measure the thermal conductivity of highly porous dust aggregates and obtained very low value of 0.002 – 0.02 W m$^{-1}$ K$^{-1}$. Assuming a nucleus with a density of 532 kg m$^{-3}$ (Jorda et al. 2016) and a heat capacity of 770 J kg$^{-1}$ K$^{-1}$ (dust at 300 K; Winter and Saari 1969), the thermal conductivity range of 0.002 – 0.6 W m$^{-1}$ K$^{-1}$ translates into a thermal inertia in the range 28 – 495 J K$^{-1}$ m$^{-2}$ s$^{-1/2}$. Models also indicate that the thermal conductivity decreases with increasing porosity, with several orders of magnitude difference between low (<20%) and high (>70%) porosity media (Prialnik et al. 2004, and references therein). While the above methods do not provide a direct constraint for the thermal inertia of the nucleus, they demonstrate however that a low thermal inertia is fully compatible with both experimental and theoretical studies on porous aggregates, from which comets are most likely made of.

The most recent estimates are provided by the spatially resolved data of the Deep Impact spacecraft, with the temperature measurements of comets 9P and 103P. From the observed temperature map of Fig. 2, Davidsson et al. (2013)

estimated the thermal inertia of 9P from less than 50 J K$^{-1}$ m$^{-2}$ s$^{-1/2}$ in rough pitted terrains to 200 J K$^{-1}$ m$^{-2}$ s$^{-1/2}$ in smooth terrains, while Groussin et al. (2013) estimated the thermal inertia of 103P to less than 250 J K$^{-1}$ m$^{-2}$ s$^{-1/2}$. The lack of temporal coverage (i.e. flyby) is the main limitation for more accurate estimates.

**2.2.2. Progress during Rosetta**

*2.2.2.1. VIRTIS thermal inertia measurements*

The huge amount of infrared spectra collected by the -M and -H channels of the VIRTIS instruments over the full Rosetta mission, has been used to investigate the surface thermal inertia. As VIRTIS was sensitive from 0.25 to 5 micron, the thermal radiation measured comes essentially from the first layers of the regolith (in principle, one millimetre maximum). Thus any VIRTIS temperature measurement can be associated to the surface thermal properties themselves.

One big advantage of the VIRTIS Rosetta observations over the previous space missions dedicated to comets studies is the large range of phases angles and local times covered by VIRTIS thanks to the complex orbit of Rosetta. This allows disentangling the surface thermal inertia effects from the surface roughness effects. The thermal emission of a rough surface is not isotropic (not Lambertian) and the same area observed at different phases angles presents different thermal observed fluxes that can be misinterpreted as a variation of the thermal conductivity. At very short wavelengths, typically the sensitivity range of VIRTIS, this effect is critical and affects significantly the Wien's part of the Planck function.

The surface thermal inertia was estimated across the whole surface using VIRTIS data acquired between September 2014 and December 2014, between 3.5 and 3.0 au heliocentric distance, when the activity was still relatively low. At that time, most of the Northern hemisphere was illuminated by the Sun while the Southern part was experiencing winter polar night. The complex geometry of the nucleus required considering topographic effects, which include mutual shadowing, mutual heating, spin rate changes over time. Synthetic thermal spectra were generated assuming heat transfer through the subsurface by conduction, and including some roughness properties both at the nucleus shape model resolution scale (few meters) and at sub-pixel scale (Kuehrt et al. 1992). For about 20 geomorphological regions, 200 spectra were randomly selected and compared to the synthetic ones.

In general, the thermal inertia remains in the range 10 – 170 J K$^{-1}$ m$^{-2}$ s$^{-1/2}$ (Leyrat et al. 2015), similar to values found on other comets (see Sect. 2.1.1). The surface temperature increases in average by 0.5 K min$^{-1}$ just after being illuminated by the Sun. Interestingly, two groups of surfaces were identified: the smooth terrains (dusty units like Imhotep, Hatmehit, etc...) present very low thermal inertia (<30 J K$^{-1}$ m$^{-2}$ s$^{-1/2}$) while rough consolidated terrains with apparent fractures seem to conduct heat more efficiently, with higher thermal inertia values (>110 J K$^{-1}$ m$^{-2}$ s$^{-1/2}$). Since the thermal inertia depends on density and porosity (e.g. for the thermal conductivity), the change in thermal inertia could result mainly from the change in the material properties, from consolidated to unconsolidated, rather than from a change in material composition.

Despite the very smooth texture of the Hapi region, this area presents two interesting "anomalies": first, the Hapi region is about 30 to 40 K warmer than expected, and second, its thermal inertia reaches intermediate values (60 J $K^{-1}$ $m^{-2}$ $s^{-1/2}$). The high temperatures can be easily explained by the mutual heating due neighbours regions Hathor and Seth that heat Hapi even in the night. A possible explanation for the medium thermal inertia is the role played by the ice trapped inside the regolith and detected by VIRTIS (De Sanctis, 2015). Water gas has condensed near the cold surface during night, which increases the effective thermal inertia. Ice favours a better contact between grains at the surface, and thus a better heat conduction.

A clear correlation exists between the distribution of local gravitational slopes (Groussin et al. 2015) and the thermal inertia (Leyrat et al. 2015). Dusty units with very flat surfaces present in general low thermal inertia, while consolidated terrains with very steep slopes are consistent with very high thermal inertia values. This behaviour indicates that low thermal inertia areas consist of fluffy debris accumulated that come from other sources: the Southern hemisphere (Keller et al. 2017) during its very short and strong summer close to perihelion, and the ejecta of neighboring units created by crack and thermal stress (Groussin et al. 2015). On the highly inclined surfaces, the regolith cannot remain stable and it is moved away, allowing the denser sub-surfaces to be exposed directly to solar light.

More recently, a more detailed analysis was initiated using an accurate shape model of the nucleus and dividing each geomorphological area in sub-regions, in particular to better understand the variations of thermal properties within the large units (Leyrat et al. 2019). Preliminary results confirm the averages values already found for the thermal inertia, with local variations from 5 to 350 J $K^{-1}$ $m^{-2}$ $s^{-1/2}$.

*2.2.2.2. MIRO thermal inertia measurements*

As described in Sect. 2.1.2.2, MIRO is capable of measuring two subsurface temperatures from different depths. Schloerb et al. (2015) estimated that the measured sub-mm and mm temperatures arise from depths of 1 cm and 4 cm below the surface, using a model of the nucleus thermal emission that takes into account thermal inertia and absorption properties at the MIRO wavelengths (for lunar soils and cometary analogs). With the MIRO dataset, several attempts have been made to estimate the thermal inertia of the comet subsurface.

The most commonly used method to calculate the thermal inertia from MIRO measurements has two steps. First, the temperature profile in the nucleus subsurface must be found, by solving the 1D heat transfer equation for the propagation of energy into the nucleus. The temperature profiles are then found for a range of thermal inertia values. In the second step, the resulting profiles serve as the input to a radiative transfer model that computes the simulated brightness temperatures. Comparing the simulated brightness temperatures to the measured brightness temperatures from MIRO enables the best estimate for thermal inertia to be found from the range of values that went into solving the heat conduction equation. This method is used by Gulkis et al. (2015), Choukroun et al. (2015), Schloerb et al. (2015) and Marshall et al. (2018).

Using MIRO measurements from August 2014, Gulkis et al. (2015) found the thermal inertia to be low, with values in the range 10 – 50 $J\ K^{-1}\ m^{-2}\ s^{-1/2}$. Choukroun et al. (2015) analysed data from the polar night side obtained between August – October 2014 and also found low thermal inertias, 10 – 40 $J\ K^{-1}\ m^{-2}\ s^{-1/2}$ in the sub-mm channel and 20 – 60 $J\ K^{-1}\ m^{-2}\ s^{-1/2}$ in the mm channel. To explain the difference in thermal inertia between the two channels, the authors suggest that ice may be present near the surface layer, which would extend the electrical penetration depth for each channel. This brings the data and best fitting values into better agreement. Work by Schloerb et al. (2015) with data from September 2014, gives 10 – 30 $J\ K^{-1}\ m^{-2}\ s^{-1/2}$ for the thermal inertia. With a value of 22 $J\ K^{-1}\ m^{-2}\ s^{-1/2}$, they check their modelled value against diurnal phase curves for the temperature and find good agreement between the data and model. Finally, Marshall et al. (2018) also analyse data from September 2014 and calculate bounds for the thermal inertia in five spots on the nucleus surface. Results in the mm channel imply that the thermal inertia is <80 $J\ K^{-1}\ m^{-2}\ s^{-1/2}$. The sub-mm channel gives similar results to the mm channel, but also imply that the thermal inertia could be higher, although this may be an artefact due to the limited available constraints on the electrical skin depth.

*2.2.2.3. Philae thermal inertia measurements*

The first estimation of the local thermal inertia at Philae's final landing site was given by Spohn et al. (2015), based on MUPUS-TM measurements and thermophysical modelling. Without the availability of a local digital terrain model (DTM) at the time of data processing, information about illumination conditions had to be inferred from other sources. While the MUPUS-PEN did not intrude into the cometary material, above-surface measurements could be used for comparison with TM measurements. The fact that the PEN did not capture a peak in temperature, evident in TM data, led to the conclusion that only a fraction of the surface in the field-of-view of the TM was illuminated by direct sunlight for a short duration of about 40 min, whereas indirect illumination by scattered visible light and thermal re-radiation from the environment was assumed to be responsible for the observed slow increase in brightness temperature after sunrise in the Abydos region (see also the discussion in Sect. 2.1.2.3). This interpretation has later been confirmed by Kömle et al. (2017), upon the availability of a local DTM in combination with the coarser global shape model of the comet.

In their modelling, Spohn et al. (2015) further assumed that the indirect fraction of the total illumination was proportional to the solar illumination of the anticipated landing region, which was modelled by the orientation of a single facet of the global shape model SHAP4 (Jorda et al. 2016) available at that time. Using these assumptions, Spohn et al. (2015) inferred a local thermal inertia at Abydos of 85 ± 35 $J\ K^{-1}\ m^{-2}\ s^{-1/2}$. Unfortunately, the derived temperature curve used for this analysis suffers from an error in the conversion between measured TM brightness temperature and kinetic surface temperature as already discussed in Sect. 2.1.2.3. This led to a distortion of the diurnal temperature curve, more pronounced in the low temperature regime. Therefore, the corrected temperatures shown in Fig. 5 are now about 8 – 10 K higher for the lowest temperatures encountered, whereas the difference is only about 2 K for the peak temperatures of approximately 130 K. This flattening of the temperature curve results in a higher thermal inertia. By comparison with the temperature curve and the error bars in Fig. 2 of Spohn et al. (2015), one gets a corrected thermal inertia of at least 120 J $K^{-1}$ $m^{-2}$ $s^{-1/2}$. Assuming that the surface is a dust/ice mixture with a density of 500 kg $m^{-3}$ and a heat capacity of <540 J $kg^{-1}$ $K^{-1}$ (corresponding to a dust/ice mass ratio >2 at 120 K, see Winter and Saari (1969), Herman and Weissman (1987)), this value for the thermal inertia implies a thermal conductivity >0.05 W $m^{-1}$ $K^{-1}$.

The local value determined at Abydos seems compatible with those deduced from remote sensing observations (Sect. 2.2.2.1 and Sect. 2.2.2.2) though it is more on the high side of all derived values. In this respect it could play a role that Abydos lies on the consolidated region of Wosret and is part of a rough talus with numerous boulders (Poulet at al., 2016). The landing site seems to be free of dust (at least down to the 1 mm resolution of the cameras), which could mean that the determined thermal inertia value is representative of that of boulders. Poulet et al. (2016) also found that the texture at Abydos falls into two classes, pebbles in the size range 3 – 6 mm, and rough material (unresolved, size <1 mm) in between. These findings fit well with the updated lower bound for the thermal conductivity derived from the MUPUS-TM data. Applying the theory of Sakatani et al. (2017, 2018) for the thermal conductivity of powdered material, it is found that for the very low temperatures encountered at Abydos, grain radii <1 cm, and reasonable comet porosities >70%, the surface energy driven adhesion and radiation are not sufficient to explain the effective thermal conductivity >0.05 W $m^{-1}$ $K^{-1}$. Some sort of cementation or sintering seems to be required to explain the relatively high thermal conductivity derived for Abydos.

It should be noted that the derivation of the Abydos thermal inertia described above is based on rather simplifying assumptions, the strongest one being the assumption that the indirect energy input is directly proportional to the average solar illumination of the environment in Abydos. This may be an oversimplification of the complex local topography, and, furthermore, it requires a low thermal inertia for the surroundings to be realistic (since the method implicitly assumes that not only the visible but also the infrared energy input to the TM measurement spot ceases after sunset). Attempts to apply more realistic and sophisticated models to the problem were performed more recently. Kömle et al. (2017) investigated the illumination and thermal environment at Abydos using a 3D thermal model and geometry based on a combination of (at that time) available global and local digital terrain models. Pelivan (2018) coupled a 1D thermal model with a high-resolution geometrical model and applied it to the Abydos region. These two models can reproduce some of the observed features like the approximate length of illumination in the Abydos region and the duration of the direct illumination of the TM spot but so far the measurements could not be reproduced quantitatively. This can mostly be attributed to the imperfections in the detailed local geometry of the complex topography at Abydos that was used in these works.

To conclude, the value of the thermal inertia at Abydos is not yet finally determined, and more modelling work should be done to derive accurate thermal properties at Abydos in the future, in particular if better local DTMs become available. Furthermore, in future models, the constraints from all Philae data should also be taken into account. This comprises the available imaging and DTM information from CIVA and ROLIS cameras, but also the housekeeping data from the lander (Sect. 2.1.2.3).

## 2.3. Synthesis on nucleus thermal properties

The thermal properties of cometary nuclei are characterized by a thermal inertia in the range $0 - 350$ J K$^{-1}$ m$^{-2}$ s$^{-1/2}$. When the nucleus is unresolved, the thermal inertia is estimated to less than 50 J K$^{-1}$ m$^{-2}$ s$^{-1/2}$, which corresponds to a low thermal conductivity of $0 - 0.006$ W m$^{-1}$ K$^{-1}$ assuming a nucleus with a density of 532 kg m$^{-3}$ (Jorda et al. 2016) and a heat capacity of 770 J kg$^{-1}$ K$^{-1}$ (dust at 300 K; Winter and Saari 1969). Spatially resolved observations show variations across the surface, depending on the nature of the terrain: on 67P, smooth terrains covered by deposits usually have a lower thermal inertia (<30 J K$^{-1}$ m$^{-2}$ s$^{-1/2}$) than terrains exposing consolidated materials (>110 J K$^{-1}$ m$^{-2}$ s$^{-1/2}$). As illustrated by Fig. 6, comets are among the bodies of the solar system with the lowest thermal inertia, taking into account the fact that the thermal inertia is a function of temperature and therefore decreases with heliocentric distance $r_h$ for a given body (Rozitis et al. 2018) with a power law of $r_h^{-0.75}$ (Delbó et al. 2007).

The values of thermal inertia derived from MIRO data are globally lower than those derived from VIRTIS (Sect. 2.2.2.1) and MUPUS (Sect. 2.2.2.3) data. This apparent discrepancy could be real, in which case there would be a decrease of the thermal inertia with depth in the first centimetres. The discrepancy may also result from the thermal models used by the different authors, which neglect the sublimation of ices and the cooling by outgassing (i.e., they use "asteroid-like" thermal models) and therefore overestimate the temperature inside the nucleus, resulting in an underestimation of the thermal inertia. This effect is more pronounced for MIRO than for VIRTIS and MUPUS, since MIRO probes greater depths. Finally, it is worth mentioning that thermal inertia is temperature dependant and is therefore expected to vary diurnally and seasonally across the surface for a given area.

Overall, a thermal inertia of $0 - 60$ J K$^{-1}$ m$^{-2}$ s$^{-1/2}$ is probably the best estimate for the bulk value, which corresponds to a low thermal conductivity of $0 - 0.009$ W m$^{-1}$ K$^{-1}$ with our previous assumptions on density and heat capacity. This bulk value is consistent with the range of $0 - 50$ J K$^{-1}$ m$^{-2}$ s$^{-1/2}$ derived from spatially unresolved observations, which by definition is an average over the nucleus surface, and moreover comes from a large sample of ~60 cometary nuclei (see references in Table 1). This bulk value is also consistent with the range of $10 - 60$ J K$^{-1}$ m$^{-2}$ s$^{-1/2}$ derived from MIRO observations of the night side of 67P (Choukroun et al. 2015); we consider this estimate as particularly reliable since the nucleus is in a purely cooling conductive regime at that time (no sun light), which simplifies the number of hypothesis in the analysis (e.g. no issues with solar insulation, roughness, surface reflection and multiple scattering). Finally, larger values are found on exposed consolidated terrains, which are usually harder, denser and less porous than the bulk nucleus (Sect. 3 and Sect. 4).

Because of the low thermal inertia, the nucleus temperature decreases very rapidly with depth, from hundreds of Kelvin to tens of Kelvin in the first meter. In addition, the erosion rate of the nucleus is, on average, comparable to the propagation velocity of the heat wave, i.e. typically decimetres to meters at each perihelion passage (Prialnik et al. 2004; Gortsas et al. 2011; Keller et al. 2015). The low thermal inertia and rapid erosion provide a good thermal insulation of the nucleus interior, which remains unaffected by solar insulation below the first meter (Capria et al. 2017), and leads to the conclusion that comets could still hold primordial materials, unaffected since their formation billions of years ago. This is mostly relevant for small comets (diameter <10 km) formed by the slow (~3 Myr) hierarchical agglomeration of materials, likely those of Fig. 1, whose internal temperature never exceeded 100 K since their formation, even in the presence of radioactive heating, allowing them to keep their super volatiles species (e.g. CO) trapped in the amorphous water ice (Davidsson et al. 2016).

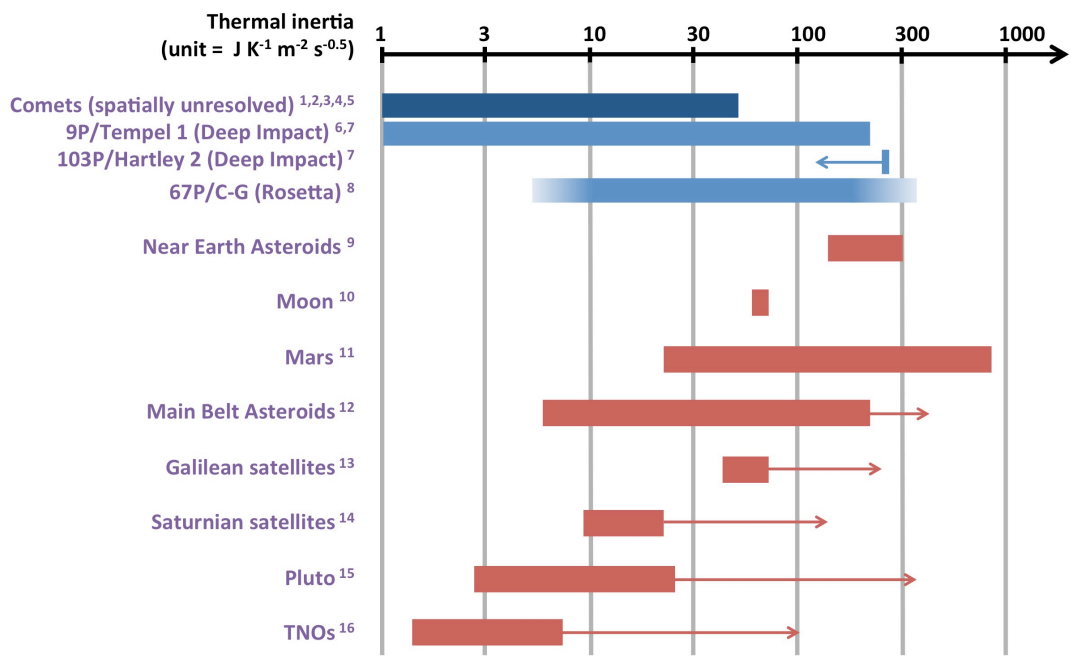

**Fig. 6** Synthesis of the thermal inertia of cometary nuclei compared to other solar system bodies. For 103P, the blue arrow indicates the upper limit (<250 J $K^{-1}$ $m^{-2}$ $s^{-1/2}$). For 67P, the blue bar indicates most estimates (10 – 170 J $K^{-1}$ $m^{-2}$ $s^{-1/2}$) with extreme values in colour gradient (<10 or >170 J $K^{-1}$ $m^{-2}$ $s^{-1/2}$). The red arrows indicate the corrected value at 1 au for bodies beyond Mars, using power law dependence with heliocentric distance $r_h^{-0.75}$ (Delbó et al. 2007). References: (1) Jorda et al. (2000), (2) Lamy et al. (2008), (3) Lisse et al. (2009), (4) Boissier et al. (2011), (5) Fernandez et al. (2013), (6) Davidsson et al. (2013), (7) Groussin et al. (2013), (8) Marshall et al. (2018), (9) Delbó et al. (2007), (10) Hayne et al. (2017), (11) Mellon et al. (2000), (12) Delbó et al. (2009), (13) Morrison and Cruikshank (1973), (14) Howett et al. (2010), (15) Lellouch et al. (2017), (16) Lellouch et al. (2013).

# 3. The nucleus mechanical properties

The mechanical properties of a comet nucleus drive its ability to respond to an internal or external stress. They are key to properly describing many processes that occurred during the history of a comet, including accretion, encounters with giant planet, or collisions. The mechanical properties can be modified by internal heating and phase transitions, and therefore also provide constraints on these processes. The mechanical properties are described by the tensile, shear and compressive strengths (Sect. 3.1), and by the elastic properties (Sect. 3.2). By definition, in solid-state physics, the tensile strength ($\sigma_T$ in Pa) is the maximal mechanical tension that can build up inside a material before it breaks, the shear strength ($\sigma_S$ in Pa) is the maximum shear load that can be applied to a material without failure, and the compressive strength ($\sigma_C$ in Pa) is the maximum load that can be applied to a material without changing the size of the sample. Usually, $\sigma_T < \sigma_S < \sigma_C$ for ductile and brittle materials.

## 3.1. Strengths

### 3.1.1. State of the art before Rosetta

Before the Rosetta mission, the tensile, shear and compressive strengths of the cometary material had been determined using observational constraints, laboratory experiments and modelling.

From the observational point of view, it is, a priori, possible to estimate the strength of the nucleus when it experiences a mechanical stress. This stress can be of various origins, including the nucleus rotation (i.e. resistance to centrifugal forces), the encounter with a giant planet or the Sun (i.e. resistance to tidal forces), or an impact (i.e. resistance to external mechanical forces). From the rotational period of several nuclei, Davidsson (2001) and Toth & Lisse (2006) estimated the nucleus tensile strength to be 1 – 100 Pa. From the close encounter of comet SL9 with Jupiter, Asphaug and Benz (1996) estimated its tensile strength to 5 Pa, while Klinger et al. (1989) estimated that of a 1 km radius sungrazing comet to 100 Pa. The Deep Impact mission, with a controlled impact of a 364 kg impactor at 10.3 km s$^{-1}$ on comet 9P, resulted in a measured shear strength of <65 Pa (A'Hearn et al. 2005). This value is however uncertain since Holsapple and Housen (2007) showed that the Deep impact experiment is compatible with any shear strength in the range 0 – 12 kPa. From all the above constraints, it however emerges the picture of a very weak nucleus, with a bulk tensile strength most probably lower than 100 Pa. As we will see in the next two paragraphs, this low strength is supported by both experimental and theoretical studies.

From laboratory experiments, it is possible to measure the strengths of cometary material analogs, with the strong limitation that we do not know well the nature of the cometary material, therefore the ground truth of the studied analogs. Various analogs made of water ice grains and/or dust grains have been studied, with an estimated tensile strength in the range 2 – 4 kPa for pure 200 micron size water ice grains (Bar-Nun et al. 2007) and in the range 1 – 10 kPa for micrometre size siliceous grains (Blum et al. 2006). The compressive strength has also been estimated to 0.3 – 1 MPa by Jessberger and Kotthaus (1989) for micron size low density aggregates of water ice and dust grains. Recent experiments are discussed in Sect. 3.1.2.3 and show that dust aggregates can have a tensile strength of only 1 Pa (e.g. Blum et al. 2014).

Theoretically, it is possible to calculate the strength of an aggregate from the van der Waals forces, depending on the physical properties of the aggregate grains (size, shape, porosity, contact area, …). Using this method, Greenberg et al. (1995) derived a tensile strength of 270 Pa for a highly porous (80%) aggregate of micron size dust/ice grains, while Kührt and Keller (1994) derived a tensile strength of ~10 kPa between two micron size grains. Finally, we can mention the work of Biele et al. (2009) who derived a tensile strength of 5 kPa for a material similar to that of Greenberg et al. (1995). More details on modelling and results from recent works are presented in Sect. 3.1.2.4.

While the bulk tensile strength of the nucleus is likely low (<100 Pa), there exist variations across the surface and with depth. The unconsolidated materials and fine deposits that cover a large fraction of the surface have, by definition, a lower strength than the consolidated materials from which, for example, boulders and cliffs are made of. Moreover, Jessberger and Kotthaus (1989), Grün et al. (1991), Thomas et al. (1994), and Pommerol et al. (2015) have shown experimentally that a hard (typically 1 MPa) layer of water ice can be produced by sublimation/redeposition cycles and/or sintering of water ice close to the surface, a result consistent with a modelling performed by Kossacki et al. (2015). This processed layer is discussed in more detail below (Sect. 3.1.2), in particular for the interpretation of the Philae measurements.

Finally, it is important to note that the strength is a scale dependent parameter, i.e. the same material has a different strength at the macroscopic and microscopic scale. The general trend is a decreasing of the strength with increasing scale, so that, for example, micron size aggregates have a larger strength than a kilometre body made of such aggregates. The scaling law is material dependent, which makes the comparison between the strengths determined from observations, laboratory experiments and modelling difficult.

### 3.1.2. Progress during Rosetta

*3.1.2.1. Strengths from remote OSIRIS observations*

Images of the nucleus of 67P acquired by the OSIRIS cameras (Keller et al. 2007) show a varied and complex morphology including surface features with sharp topography, such as pits, cliffs, overhangs and fractured consolidated material, suggestive of non-zero material strength.

Groussin et al. (2015) estimated the strength of collapsed surface features, i.e. the tensile strength that was presumably overcome by the weight of overhanging material in order to collapse, and a compressive strength scaled from this. They estimate the consolidated material to have tensile strength 3 – 15 Pa and compressive strength 30 – 150 Pa at 5 – 30 m scales. They also estimate the shear strength or cohesion needed to hold meter sized boulders on slopes as 4 – 30 Pa, and to resist the lateral pressure at the bottom of the 900 m high Hathor cliffs as >30 Pa.

Basilevsky et al. (2016) performed a similar analysis of overhangs and landslides, resulting in comparable strength estimates. Vincent et al. (2017) also measured the cohesion needed to prevent the collapse of cliffs, finding 1 – 2 Pa strengths at tens of metre scales globally. They then suggest weakening by sublimation effects dominate over collapse under self-gravity.

Attree et al. (2018a) performed a survey of overhanging cliffs, using a full 67P shape model (Preusker et al. 2017) with calculated gravity vectors to identify facets with > 90° local slope. They then examined twenty features in detail, measuring a profile of each and deriving a minimum tensile strength estimate. Strengths ranged from 0.02 – 1.02 Pa, with no apparent correlation with position on the nucleus head or body. Attree et al. (2018a) argue that the presence of debris at the base of nearly all these overhangs and the observed collapse of two, as well as their depths exceeding likely zones of thermal processing (based on thermal modelling), confirms the comet's low global bulk tensile strength.

Despite these low strengths, the prevalence of fractures on 67P indicates a material with sufficient strength or stiffness to behave in a brittle way. Fractures are present at all scales in OSIRIS images, from hundreds of metres down to tens of cm (Thomas et al. 2015; El-Maarry et al. 2015; Pajola et al. 2015; Matonti et al. 2019); and cm and below scale in images from the Philae lander (Poulet et al. 2016). They may be caused by stresses associated with rotational and shape effects, tidal forces, collisions and thermal cycling. Metre-scale polygonal fracture networks, in particular, resemble similar polygon features on Earth and Mars (Auger et al. 2018) and are probably associated with thermal stresses, as discussed below in Sect. 3.1.2.3.

Reconciling low bulk strengths and granular materials with stiff/brittle surface features is a difficult challenge that most likely has its answer in the stratigraphy and hardening and sintering processes described in the next sections. Groussin et al. (2015) also note, finally, that the ratio of the derived material strengths to surface gravity for 67P is similar to that of Earth, which might explain the perceived morphological similarities.

*3.1.2.2. Strengths from in-situ Philae measurements*

Despite the unintended excursion of Philae after its touchdown at the Agilkia landing site, the lander could contribute to the investigation of cometary strength in several ways – it was even possible to instrumentalise the bouncing itself in this regard. The associated measurements and observations occurred sequentially, thus we summarize the results in chronological order.

The images of Agilkia taken by the ROLIS camera (Mottola et al. 2007) confirmed the interpretation of earlier OSIRIS images that many subhorizontal surfaces of 67P are covered by a granular material, made of particles bonded so weakly that moat- and wind-tail-like structures can form around boulders under airfall of particles ejected from elsewhere (Mottola et al. 2015). The presence of boulders of several meters on the other hand shows that the granular medium is a mere cover.

All three accelerometers of CASSE were set to triggered mode for the landing and recorded the individual touchdowns of Philae's three feet at Agilkia. Applying Hertzian contact mechanics to the deceleration time series, a compressive strength between 3.5 kPa and 12 kPa could be derived from +Y-foot data, which hit the ground first, likely on one of the boulders at the Agilkia site (Möhlmann et al. 2018).

After touchdown, Philae bounced off and started drifting across the comet. Images taken by OSIRIS before and after the touchdown show fresh excavations up to 20 cm deep in the granular medium, resulting from the bouncing. Biele et al. (2015) find from a parameter study that the uniaxial compressive strength of the material is unlikely to exceed ~10 kPa, since otherwise the depth of the excavations would be reduced to millimetres. They also argue that a layer of much larger strength is located below these surface materials. The depth of the footprints at the same time poses a lower limit for the thickness of the granular layer. Roll and Arnold (2016) further elaborate this evaluation and find an average compressive strength of 2 kPa, or an increase of 3 kPa m$^{-1}$, both results being valid for the depth the soles penetrated into the ground.

At the final Abydos landing site, a non-granular, fractured surface was found (Bibring et al. 2015). MUPUS attempted to hammer its thermal probe, a 33 cm carbon rod with metal tip, into this material but failed to penetrate. A comparison with laboratory tests shows that an uniaxial compressive strength of the material of at least 2 MPa is required to frustrate the penetration of MUPUS (Spohn et al. 2015).

MUPUS nevertheless hammered for more than three hours, and CASSE recorded 14 of its hammer strokes (Knapmeyer et al. 2018, and references therein). The dispersion of recorded surface waves indicates a decrease in shear wave velocity with depth, suggesting that strength is depth dependent (see also Sect. 3.2).

*3.1.2.3. Strength from laboratory experiments*

Due to their formation history, cometary surfaces are believed to consist of granular material possessing very high porosities (~80 %; Sierks et al. 2015) and a composition made of ices, organic materials and minerals (Filacchione et al. 2018). Two different configurations of the cometary material as a result of the formation and evolution of the cometary nucleus are currently under debate (Fig. 7): homogenous dust layers (see, e.g., Davidsson et al. 2016) versus aggregate layers (see, e.g., Blum et al. 2017). These two different configurations have a strong influence on the strength of the surface material, such as the tensile and the compressive strength. In the past, laboratory experiments were utilized to investigate the strength of granular materials under cometary conditions.

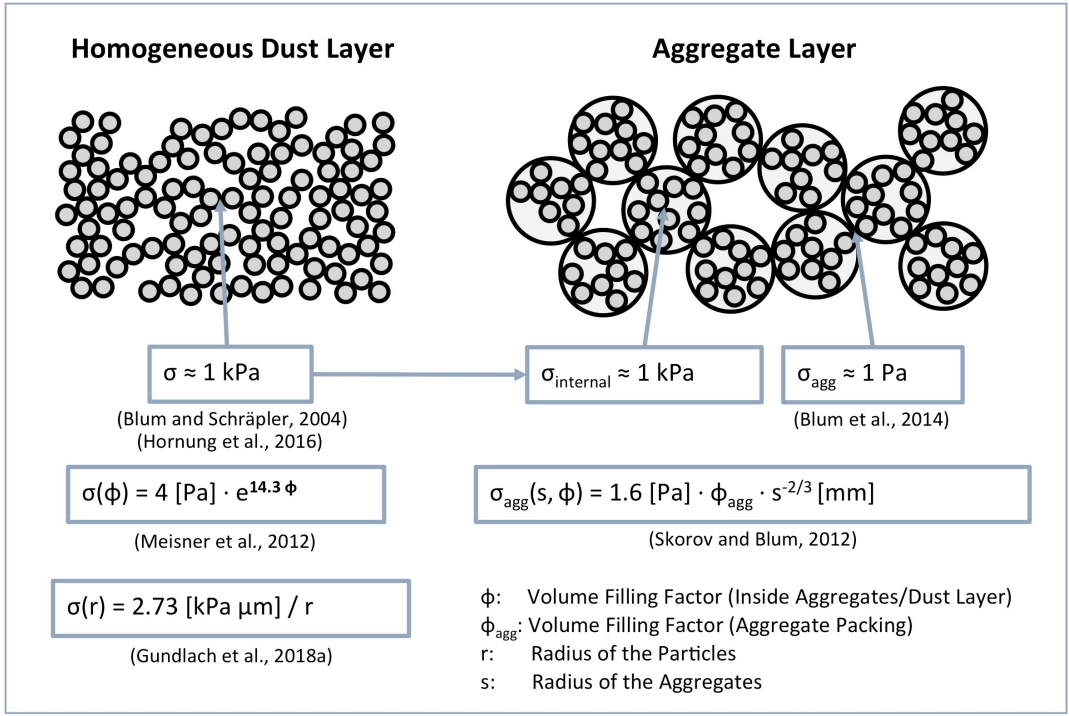

**Fig. 7** Summary of the main laboratory results (including the COSIMA findings). The corresponding references are provided below the equations.

The tensile strength of homogenous dust layers (composed of silica particles with a diameter of 1.5 μm) was measured by Blum and Schräpler (2004). They found that the dust layers (porosity ~0.8) possess a tensile strength of ~1 kPa. Meisner et al. (2012) utilized the Brazilian Disc Test to measure the tensile strength of centimetre-sized dust samples (composed of irregular silica particles in the 0.1 to 10 μm-size regime) and estimated values in the range from 1 to 5 kPa. Furthermore, a relationship between the tensile strength, $σ_T$, and the volume filling factor, $φ$, was derived, which reads, $σ_T = 4\ Pa \cdot exp(14.3\ φ)$. The same approach was used by Gundlach et al. (2018a; their Fig. 4) to study the influence of the grain size on the tensile strength of granular materials. The result was that the tensile strength is decreasing linearly with increasing grain size. Furthermore, the tensile strength of granular water ice was estimated to 0.9 ± 0.7 kPa, for samples consisting of spherical poly-disperse particles with a mean radius of 2.38 ± 1.11 μm and a porosity of 0.5. This value is ten times lower than expected based on the increased specific surface energy of water ice compared to silica (Gundlach et al. 2011). The only explanation for this behaviour is that water ice possesses a similar specific surface energy as silica at low temperatures (0.02 J m$^2$). The results of laboratory experiments investigating the tensile strength of homogenous granular materials are in good agreement with astrophysical measurements such as the breakup of cometary dust particles observed with the COSIMA experiment onboard the Rosetta spacecraft (~1 kPa; Hornung et al. 2016) and with the derived tensile strength from meteor streams breakup in the Earth atmosphere (0.4 – 150 kPa by Blum et al. 2014; 40 – 1000 Pa by Trigo-Rodríguez et al. 2006).

Thus, the tensile strength of homogenous dust layers can be used to describe the internal tensile strength of dust aggregates (sub-decimetre-sized objects composed of micrometre-sized particles). Measurements of the tensile strength of packings of aggregates have, however, revealed much lower values, in the order of only ~1 Pa (Blum et al. 2014). Furthermore, the measurements have shown that the tensile strength of aggregate packing, $\sigma_{agg}$, decreases with aggregate size, s, with the relation, $\sigma_{agg} \sim s^{-2/3}$, as predicted by Skorov and Blum (2012). It is therefore important to note that the tensile strength of a granular material can vary by at least three orders of magnitude just by different arrangements of the material (homogeneous dust layers versus aggregate packings).

In contrast, the compressive behaviour of granular materials cannot be expressed by just a single value. In solid-state physics the compressive strength is a measure for the maximum load that can be applied to a material without changing the size of the sample. However, granular materials have the tendency to react with deformation also if very small loads are applied. Thus, a better description of the behaviour of granular materials under compression is given by "compression curves". While compressed, the size change leads to an increase of the volume filling factor of the material. Figure 8 shows the compression curves of different granular materials measured in the laboratory (Güttler et al. 2009; Schräpler et al. 2015; Lorek et al. 2016). The onset of deformation and the turnover point are characteristics of S-type functions that can be used to derive material properties such as the rolling friction force and, therewith, the specific surface energy, or the particle radius (see Eqs. 5-7 in Schräpler et al. 2015). In order to derive material properties from laboratory, or spacecraft measurements it is mandatory to either derive the entire compression curve, or to estimate the change of the volume filling factor of the material by, e.g., a measurement of the intrusion depth, if only single data points can be acquired. In this context, an application to the MUPUS findings (>2 MPa; Spohn et al. 2015) to derive cometary material properties is very challenging, not mentioning that the compression curve is also scaled dependent.

Furthermore, hardening of the cometary surface layers can occur by, e.g., sintering of water ice, a process that transports molecules from the particles in contact into their neck region (Kossacki et al. 1994). The growth of the sinter neck can significantly increase the strength of the material as shown by Fig. 9 (derived from the sinter model provided by Gundlach et al. 2018b). However, it is important to note that the sintering is only a short-term effect because sublimation will always lead to evaporation of the contact area. For comets, the sinter timescale is faster than the erosion process, but the sinter neck has a lifetime of $10^{-2} - 10^3$ s, depending on the ice temperature (220 – 160 K, respectively; Fig. 12 in Gundlach et al. 2018b). Thus, the sinter neck can form in ice-rich areas at these temperatures, but the increase of the tensile and the compressive strength are only temporary.

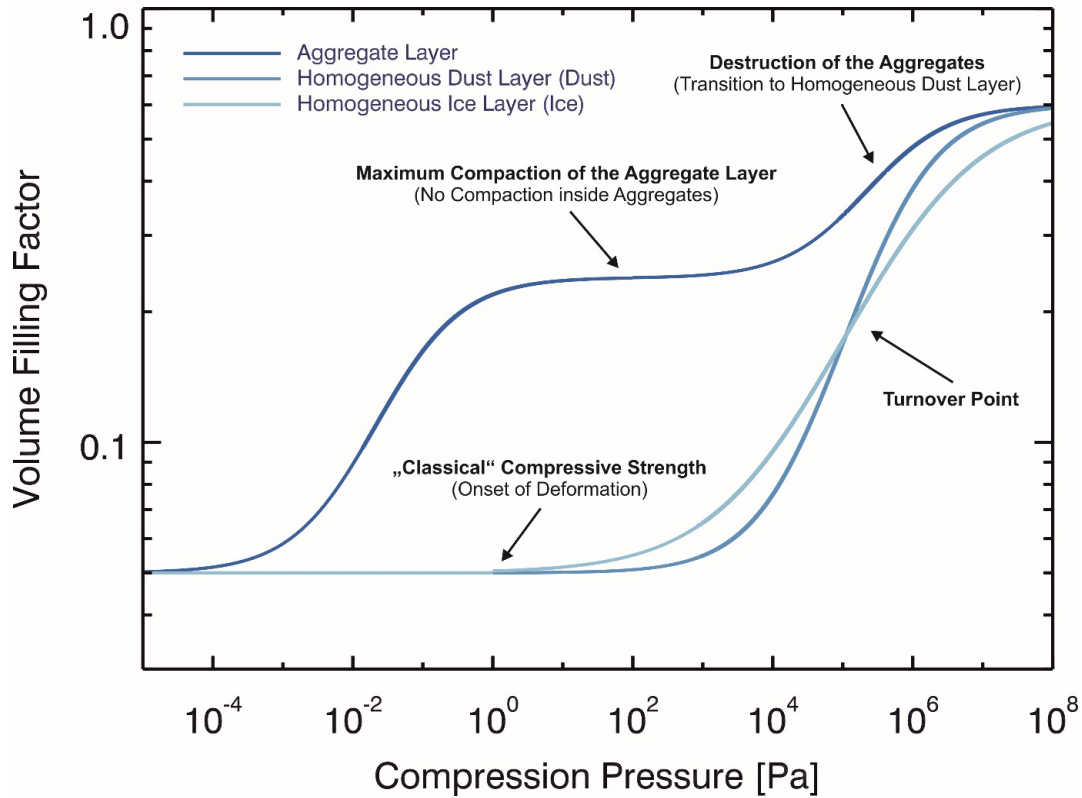

**Fig. 8** Compressive strength of granular materials measured in the laboratory: aggregate layers (Schräpler et al. 2015), homogeneous dust layers (Güttler et al. 2009) and homogeneous water-ice layers (Lorek et al. 2016). Since all the samples start with a different initial volume filling factor, the compression curves were normalized to an initial and a maximum volume filling factor of 0.05 and 0.6, respectively. The first compaction is due to the rearrangement of the structure, aggregates are rolling, etc… Later the aggregates are destroyed as indicated in the figure.

Such a hard layer should experience significant thermo-mechanical stresses due to seasonal and diurnal temperature changes, a process which may be an important source of weathering on asteroids, the moon and other airless bodies (Delbó et al. 2014). The large temperature variations experienced by cometary surfaces can induce stresses exceeding the tensile strength of solid water ice (Kuehrt 1984; Tauber and Kuehrt 1987), suggesting they are responsible for at least some of the fractures observed on the surface of 67P (Auger et al. 2018).

Attree et al. (2018b) investigated this with a thermo-viscoelastic model, based on that used for frozen Mars soil. They found the seasonal temperature cycle on 67P to induce large thermal stresses, of up to several tens of MPa, down to tens of centimetres to ~metres depths, proportional to soil thermal inertia and ice content. This assumed a relatively stiff ice-bonded layer (see Sect. 3.2, below, for a discussion of elastic properties) but is entirely consistent with the observed ~metre-scale thermal contraction crack polygons (Auger et al. 2018). These polygons are detected at all latitudes, suggesting that a sufficiently hard layer is globally present. Thermal fracturing should, therefore, be an important erosion mechanism, contributing to the breakdown of boulders, consolidated material and cliffs on cometary surfaces.

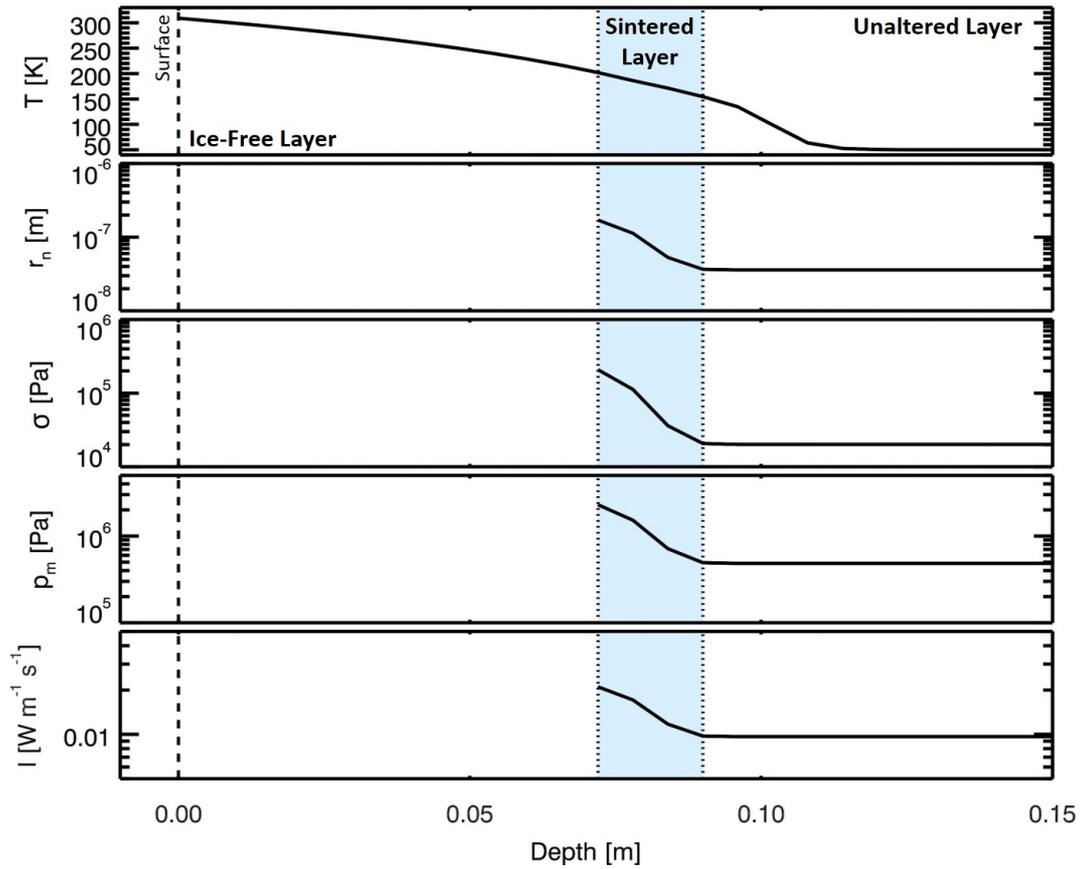

**Fig. 9** Hardening of the cometary surface by sintering of micrometre-sized water-ice particles. The first panel shows a typical temperature distribution (T) inside the surface layers of the cometary nucleus derived from a thermophysical model developed by Blum et al. (2017). In the second panel, the evolved sinter neck ($r_n$) between the ice particles is shown (based on the sinter model described in Gundlach et al. 2018b). The next three panels are visualizing the change of the tensile strength (σ), the change of the turnover point of the compression curve ($p_m$) and the change of the thermal conductivity (l).

*3.1.2.4. Strength from modelling*

Estimates of the strength of cometary material are inherently related to the evaluation of its cohesion. Since the comet nucleus undoubtedly has a high porosity, plausible estimates of cohesion can be obtained on the basis of the well-known theory (Rumpf 1962, 1970), where the tensile strength is calculated via the cohesion expressed by forces acting at the contact between the neighbouring particles. These forces may form with and without material bridges. In the first case the Van der Waals or electrostatic forces cause the adhesion. In the second case the adhesion arises from the presence of different bridges (e.g. sintering). Speaking of particle-surface contacts and their force-response behaviour one can select the following basic approaches: 1) elastic contact deformation, 2) plastic contact deformation with adhesion, 3) visco-elastic and visco-plastic contact deformations. For cometary conditions, only the first approach seems to be relevant. The theoretical basis of this approach was developed by Hertz (1882) and developed in Derjaguin et al. (1975) and Johnson et al. (1971).

For a homogeneous random porous media composed by spherical grains the tensile strength $\sigma_T$ of a porous bed can be expressed as (Rumpf 1970):

$$\sigma_T = \frac{(1-\varphi)\, N_c F_c}{4\,\pi\, R^2} \qquad (1)$$

where $\varphi$, $N_c$, $R$ and $F_c$ are the porosity, the average coordination number, the effective particle radius (for two spheres with radii $R_1$ and $R_2$ $1/R = 1/R_1 + 1/R_2$) and the adhesion force between particles. We note that without a significant loss of accuracy, one can assume that $\varphi\, N_c \approx \pi$ (Rumpf 1974, Rumpf 1990), so:

$$\sigma_T = \frac{(1-\varphi)/\varphi\, F_c}{4\, R^2} \qquad (2)$$

The adhesive force between two spheres is directly proportional to the specific surface energy $\gamma$ and $R$. For the elastic spheres two well-known expressions are usually applied:

- the so called JKR model (Johnson, Kendall, and Roberts 1971):

$$F_c = 3\,\pi\, R\, \gamma \qquad (3)$$

- and the DMT model (Derjaguin, Muller, and Toporov 1975):

$$F_c = 4\,\pi\, R\, \gamma \qquad (4)$$

As it was shown by Tabor (1977), the JKR model is appropriate for relatively large soft particles with high surface energies, whereas the DMT model works well for small hard particles with low surface energies. For our purposes, the difference between these expressions is unimportant. We have note that both JKR and DMT models are limited to spheres with a smooth surface. From the above it is evident that tensile strength is inversely proportional to the effective particle size. Thus, for the porous medium consisting of mono-disperse micron size spheres the tensile strength is about several kPa, however, for the millimetre-sized particles the tensile strength is about few Pascal only. We remind that it only considers Van der Waals forces (no sintering).

The generalization of this conclusion to a poly-disperse medium is nontrivial. It is well known that a substantial additive of fine particles can increase effective medium cohesion. At the same time, one can imagine a situation (for example, for a bi-disperse mixture) where the number of contacts per square meter is controlled by large particles, while the strength of individual contact between particles is determined by small ones. In this case, small grains perform the same role as the surface roughness of large particles, i.e. they reduce the strength.

An interesting attempt to estimate the effective strength of a hierarchical porous medium was made in Skorov and Blum (2012). Assuming that cometesimals formed by gravitational instability of a cloud of porous dust and ice aggregates in a gentle environment, the authors presented a consistent model of a porous medium having hierarchical structure. Using theoretical estimates and experimental results, they found that the effective tensile strength $\sigma_{eff}$ of the dust layer could be expressed as:

$$\sigma_{eff} = \sigma_a\,(1-\varphi)\frac{A}{A_0} \tag{5}$$

with $A$ and $A_0$ being the contact area between two dust aggregates and the cross-sectional area of the aggregates, respectively. The intrinsic tensile strength of the aggregates $\sigma_a$ is estimated by using Eq. (8) of Güttler et al. (2009). Finally, the effective strength is expressed via

$$\sigma_{eff} = \sigma_l\,(1-\varphi)\left(\frac{R}{1mm}\right)^{-2/3} \tag{6}$$

with $\sigma_l$ = 1.6 Pa a numerical factor obtained for the specific values of the model parameters given in Skorov and Blum (2012). In this model the effective strength is a function of surface energy also (this characteristics is hidden in $\sigma_l$). The predicted theoretical tensile strength of aggregates is in good agreement with the recent experimental results (Blum et al. 2014; Brisset et al. 2016). It was shown that the hierarchical structure of the surface layer formed by porous dust aggregates dramatically reduces the tensile strength of the layer to only a few Pascal.

## 3.2. Elastic properties

### 3.2.1. State of the art before Rosetta

Tidal breakup of cometary nuclei requires the tidal stress resulting from close encounters with other bodies to exceed the tensile strength of the nucleus (Sect. 3.1.1). This stress depends not only on masses, distances and dimensions of the two bodies, but, as Greenberg et al. (1995) pointed out, also on the Poisson ratio of the cometary material, although the dependency is weak.

Of greater interest is the possibility to explore the subsurface structure and stratification of the comet using the propagation of elastic waves. The CASSE instrument of Philae (Comet Acoustic Surface Sounding Experiment, Seidensticker et al. 2007) was actually designed to this end. It is expected that the elastic moduli of cometary material depend on the degree of sintering and will thus allow investigating the formation of layering predicted by several authors in the past as explained earlier (Sect. 3.1).

Before Rosetta, little was known about the elastic properties of cometary materials, while several models for composition and structure, especially layering were published. Material properties had to be derived from assumed compositions.

Stöffler et al. (1991) describe the composition and preparation of a cometary analog material that was used in the KOSI experiments at DLR Cologne in the early 1990s. Estimations of elastic moduli or elastic wave velocities can be derived from the composition using heuristic approaches based on snow, or using mixture theories. It must however be noted that the elastic moduli of granular material are dominantly defined by grain contacts rather than by grain composition (e.g. Mavko et al. 2009). A common property of granular media is indeed very low seismic velocities, even below the speed of sound in air. Shear wave velocities found in lunar regolith vary over a wider range than those in snow, and actually values as low as 30 m/s are even possible (Knapmeyer et al. 2018; their Fig. 2).

The formation of layers can be considered and is linked directly to the idea that comets are a mixture of volatile and non-volatile constituents. Whipple (1950) not only suggested the dirty snowball concept, he also concluded from this concept that a size-dependent ejection of particles, and the different sublimation properties of $H_2O$ and $CO_2$, may result in the formation of subsurface layering. The observation that most of 1P surface is inactive gave rise to models for the formation of inert crusts (see Kührt and Keller 1994 for the post-Halley state of the discussion). Kührt and Keller (1994) also discuss the importance of cohesive forces between particles, in addition to sintering and cementing. The time necessary to produce sinter bridges of a significant volume between ice grains is strongly dependent on particle size. Thus, subsurface structure is the result of interplay between accelerated sintering of surface materials, but also abrasion close to perihelion, the inward propagation of sublimation fronts of different volatiles, and airfall of ejected dust. Significant discrepancies between proposed layer thicknesses (e.g. Prialnik and Mekler 1991 versus Kührt and Keller 1994) or ablation rates (e.g. Keller et al. 2015 versus Brugger et al. 2016) indicate that ground truth even from a single point would be a valuable constraint to improve the existing models.

### 3.2.2. Progress during Rosetta

Even Philae's brief contact with the surface of 67P at Agilkia allowed a determination of the material's Young's modulus. The six cup-shaped soles of the three feet of Philae are capable of several modes of free resonances. When in contact with an elastic surface, the coupled oscillator shows modified resonances from which the Young's modulus of the surface can be derived. Using this approach, Möhlmann et al. (2018) found that the material in contact with Philae had a Young's modulus of a few MPa. This is much lower than the 10 GPa found in bubble-free water ice, but compares well to snow with porosity beyond 0.75 (Knapmeyer et al. 2018, and references therein).

The penetration phase of MUPUS at the Abydos site provided an opportunity for further analysis (Knapmeyer et al. 2018). A comparison of recorded CASSE accelerometer signals from the different feet of Philae, as well as comparison with laboratory experiments, indicates that all three feet were in contact with the comet at least during some of MUPUS hammer strokes. Several strokes were recorded on more than one foot, allowing travel times to be compared. Since the topography at Abydos is not known with sufficient accuracy (ROLIS descent images were shot at Agilkia, not Abydos), only an interval of possible propagation velocities (and thus shear modulus) could be derived. In terms of the Young's modulus, 7.2 MPa < $E$ < 980 MPa was found by Knapmeyer et al. (2018) with the lower bound being a firm limit, while the upper bound generously accounts for uncertainties in the path length. Typical correlations between Young's modulus ($E$) and compressive strength ($\sigma_c$) indicate a ratio $E / \sigma_c$ ~ 100 – 400 for ice or dust/ice mixtures (Jessberger and Kotthaus (1989), Möhlmann et al. (2018) and references therein), which translates to a compressive strength of 18 kPa < $\sigma_c$ < 9800 kPa. Finally, assuming a typical ratio $\sigma_c / \sigma_T$ ~ 10 for brittle material, it gives a tensile strength in the range 1.8 – 980 kPa for the layer that stopped MUPUS.

An independent result is the layering indicated by the frequency dependency of propagation velocity. The presence of dispersion can be established without any assumptions. Surface waves at frequencies below 250 Hz are delayed by about 10 ms with respect to higher frequency waves. Given the short distances between MUPUS and the three feet, this amounts to a velocity reduction of at least 32%. Since the penetration depth of surface waves depends on wavelength, velocity reduction and estimated wavelengths mean that the above Young's modulus applies to a top layer with a thickness of 0.1 m to 0.5 m, while the material below shows reduced Rayleigh wave velocities, probably due to a reduced Young's modulus (density variations might also contribute, but are less plausible, given the available material).

MUPUS non-penetration and thermal inertia measurements (Sect. 2.2.2.3; Spohn et al. 2015) suggest that MUPUS was stopped on a hard layer, where the Young's modulus is likely increased as well. Given the uncertainty in the CASSE Young's modulus and the MUPUS compressive strength estimates, it cannot be excluded that the fast near surface layer found by CASSE is indeed the sintered layer that stopped MUPUS PEN.

In summary, the combination of MUPUS and CASSE measurements indicates the presence of two or three subsurface layers, with a Young's modulus that decreases with depth.

### 3.3. Synthesis on mechanical properties

The mechanical properties of cometary nuclei are characterized by a low bulk tensile strength of typically 100 Pa or less, with spatial variations across the nucleus, and locally the presence of a shallow (10's of cm) sub-surface layer of harder materials with a Young's modulus in the range 7.2 – 980 MPa, i.e. a tensile strength of typically 1.8 – 980 kPa. To put these numbers in perspective, the harder material compares well to porous (75%) snow (Knapmeyer et al. 2018) or silica aerogel, while the bulk nucleus material ($\sigma_T$ of 100 Pa) has no analog on Earth among consolidated materials and only exists on comets because of the very low gravity (typically $10^{-4}$ m s$^{-2}$).

Figure 10 shows a synthesis of the strength values for cometary nuclei measured by different techniques. As explained in Sect. 3.1.1, the strength is a scale dependent parameter, which makes the comparison between the different measurements difficult and somewhat speculative. As an indication for Fig. 10, we used the scaling law of water ice, which is a power law $d^{-q}$, where $d$ is the scale and $q\sim0.6$ for water ice (Petrovic 2003). This scaling law should be taken with caution, since it is material dependent (Bažant 1999). The strength is also temperature dependent, for example water ice is 3 to 4 times stronger at 120 K than at 220 K (Schulson and Duval 2009), but this is of second order effect compared the scale changes of Fig. 10. From Fig. 10, bearing in mind the above limitations on scaling laws, we see that three out of four strengths values are <100 Pa at the meter scale. Only three estimates are incompatible with a tensile strength <100 Pa, including two from Philae surface measurements and resulting from the presence of a processed layer below the surface (Sect. 3.1 and Sect. 3.2).

As shown in Fig. 11, a bulk tensile strength of only 15 Pa is sufficient to keep comets stable against rotational splitting, under the assumption that they have densities and shapes similar to 67P. The above value of 100 Pa for the tensile strength of cometary nuclei likely provides sufficient margins against density and shape variations among comets.

Finally, the low bulk tensile strength of cometary nuclei provides an important constraint for their formation and evolution. In a first scenario, one can argue that to keep such a low tensile strength over 4.6 Gy, the formation and evolution must be gentle, favouring a primordial rubble pile origin (Davidsson et al. 2016), with a formation by low velocity (1 m s$^{-1}$) accretion of pebbles (Blum et al. 2014) for a resulting tensile strength of typically 1 Pa (Skorov and Blum 2012), rather than a collisional scenario between two 50 - 100 km size bodies at ~0.5 km s$^{-1}$ (Davis and Farinella 1997). In a second scenario, one can argue that the low bulk tensile strength is compatible with comets being collisional fragments of larger bodies (Morbidelli and Rickman 2015), in which case the nucleus could be made of numerous mechanically strong (kPa – MPa) blocks and boulders, assembled together for a resulting low bulk tensile strength (Pa), similar to rubble piles asteroids (Sánchez and Scheeres 2014). Disentangling between these two scenarios certainly requires more constraints than the strength itself, but it emphasizes the importance of understanding the scale dependence of the strength for cometary materials, from the microscopic to the macroscopic scale.

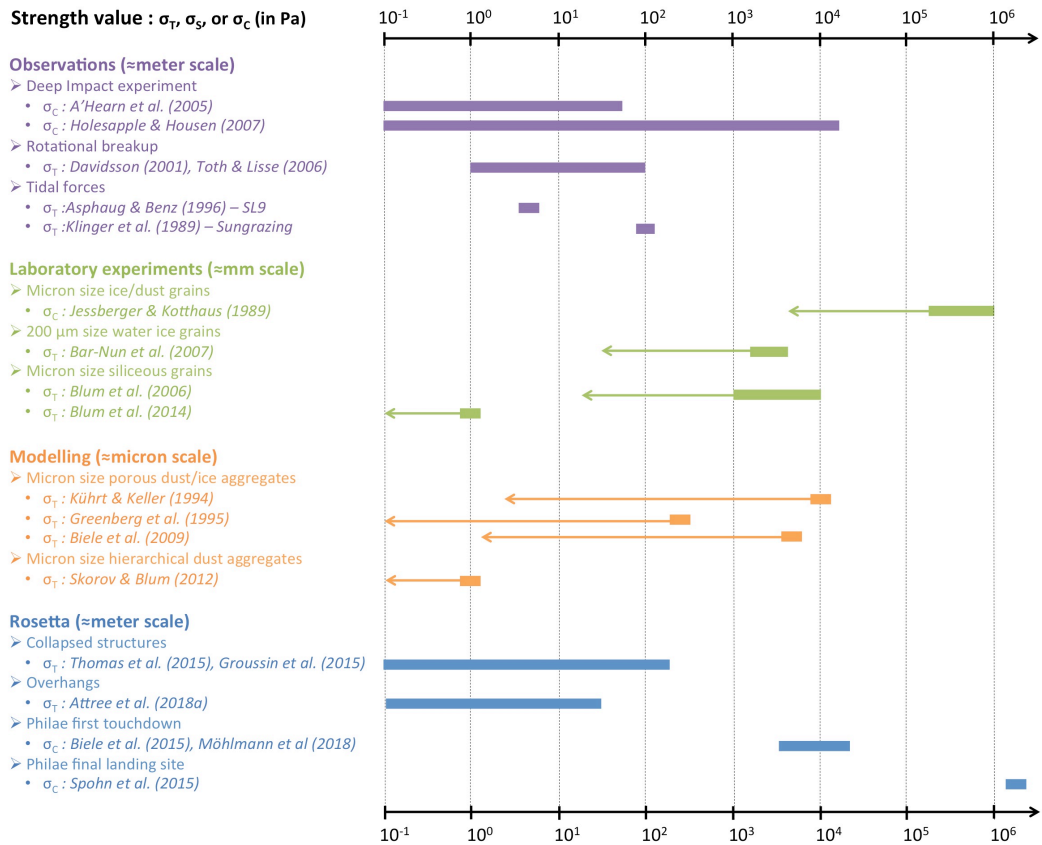

**Fig. 10** Synthesis of strength values for cometary nuclei measured by different techniques and authors, including Rosetta estimates (non exhaustive list). For each strength value, we specify whether it is the tensile ($\sigma_T$), shear ($\sigma_S$) or compressive ($\sigma_C$) strength. For laboratory experiments and modelling, the arrows indicate the strength scaled to the meter scale, using a power law with a power exponent of -0.6 (see text for details). This scaling law should be taken with caution, since it is material dependent (Bažant 1999).

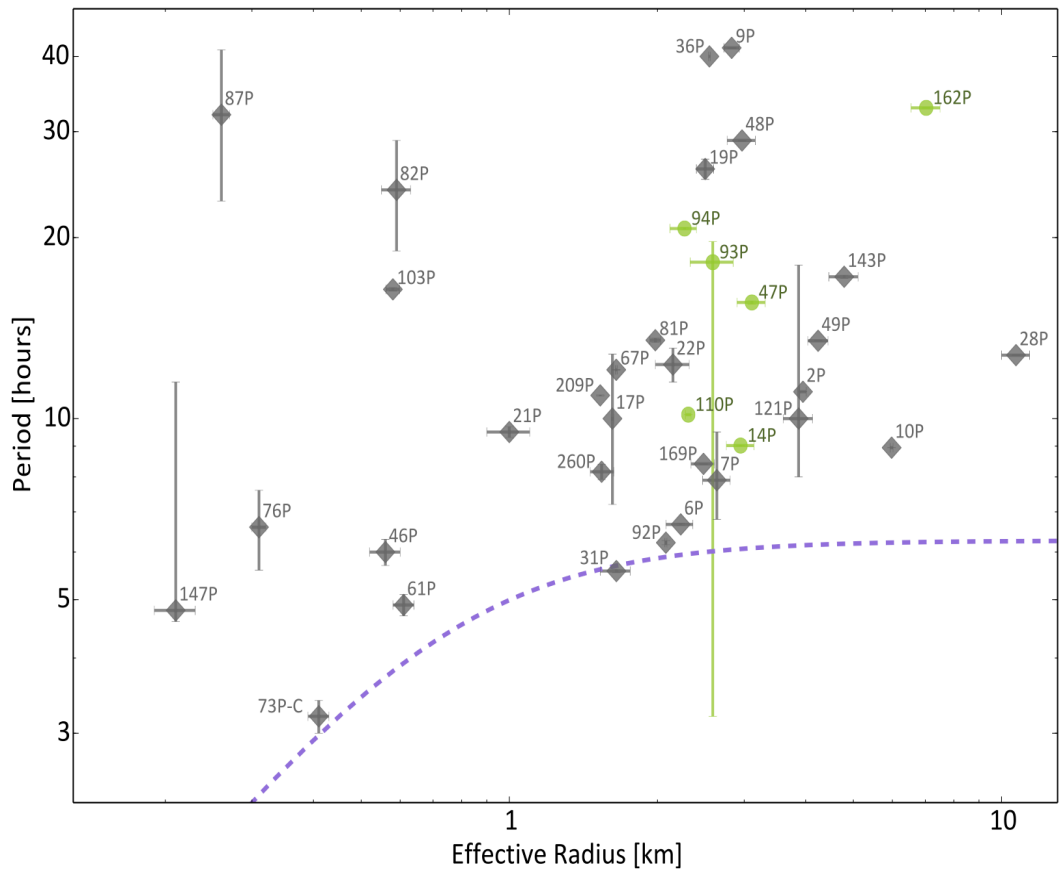

**Fig. 11** Rotation period against effective radius for JFC nuclei (adapted from Kokotanekova et al. 2017). The grey diamonds correspond to comets observed with ground- and space-based telescopes. The green circles denote the comets visited by spacecraft. The dashed curve corresponds to the model for prolate ellipsoids stable against rotational instability by Davidsson (2001), plotted for the parameters derived for 67P, i.e. a density of 532 kg m$^{-3}$, an axis ratio a/b = 2 and a bulk tensile strength of 15 Pa.

## 4. The nucleus structural and dielectric properties

The structural and dielectric properties of the comet nucleus provide information on the nature of the cometary material and in particular its roughness (Sect. 4.1), density (Sect. 4.2), permittivity and porosity (Sect. 4.3). Excepted for surface roughness, these properties are best estimated by radar and permittivity probes, which operate in the radio wavelength range, therefore providing information on the interior of the nucleus, inaccessible to visible or infrared instruments observing only the surface or sub-surface. Overall, they help to constrain and understand the nucleus vertical stratigraphy from its surface to its centre (Sect. 4.4).

### 4.1. Surface roughness

Comet nucleus surfaces are rough on different size scales. For practical reasons we here distinguish between two size regimes: 1) *large-scale roughness* (>10 m) is resolved by cameras on orbiting or flyby spacecraft, while 2) *small-scale roughness* (<10 m) is often only detectable through its shadowing effects on reflected or emitted radiation.

**4.1.1. State of the art before Rosetta**

Pre-Rosetta spacecraft imaging established the existence of rough terrain containing pits and circular depressions with a cumulative size distribution power law index $p$=-1.7 for comet 81P and $p$=-2.1 for 9P (Thomas et al. 2013a), boulders with $p$=-2.7 on 103P (Pajola et al. 2016), and smooth terrain proposed to be flows on 9P (Thomas et al. 2013a) or fallback on 103P (Thomas et al. 2013b).

Small-scale roughness obtained from optical photometric phase functions and measured in terms of the Hapke (1984) mean slope angle $\theta$ yielded $\theta$ = 13 – 25° for 19P (Buratti et al. 2004), $\theta$ = 16 ± 8° for 9P (Li et al. 2007b), $\theta$ = 27 ± 5° for 81P (Li et al. 2009), and $\theta$ = 15 ± 10° for 103P (Li et al. 2013).

Analysis of infrared radiation, that allows for the simultaneous determination of thermal inertia and small-scale roughness yielded $\theta$ = 45° for rough terrain on 9P (Davidsson et al. 2013). Roughness effects are also evident in infrared spectra of 103P (Groussin et al. 2013).

**4.1.2. Progress during Rosetta**

The unprecedented resolution and complete nucleus coverage provided by the OSIRIS cameras has allowed for the construction of accurate geometric shape models for 67P (Jorda et al. 2016; Preusker et al. 2015, 2017). Combined with methods for accurately calculating the gravity field (Werner and Scheeres 1997), this has allowed for the identification of cliffs, that is one manifestation of large-scale roughness. Vincent et al. (2017) studied the cumulative cliff height distribution and found that regions subjected to a higher level of insolation and sublimation-driven erosion have more negative $p$-values. They therefore suggest that 81P ($p$=-1.7) and 67P ($p$=-1.69 on average) have more pristine surfaces than 9P ($p$=-2.1) and 103P ($p$=-2.7). This exemplifies the importance of measuring comet nucleus surface roughness in order to better understand comet evolution.

Small-scale roughness from optical photometry yielded $\theta$ = 22 ± 7° (Fornasier et al. 2015) and $\theta$ = 19 (+4/-9)° (Ciarniello et al. 2015) for 67P. Analysis of roughness derived from infrared emission is on-going (e.g., Leyrat et al. 2014; Marshall et al. 2018). Images with extremely high resolution obtained by CIVA (1 mm px$^{-1}$) of consolidated terrain at Abydos (Bibring et al. 2015), and of smooth terrain by ROLIS (1 cm px$^{-1}$) at Agilkia (Mottola et al. 2015) and by OSIRIS (1.4 cm px$^{-1}$) at Sais (Pajola et al. 2017) have provided rare glimpses of the true nature of small-scale roughness. Smooth terrains consist of facetted and angularly shaped pebbles with sizes in the cm-dm range. Consolidated terrains, on the other hand, consist of a mixture of cemented/sintered agglomerates of mm-cm sized grains, and surfaces of fine-grained material that are smooth on decametric size scales (Bibring et al. 2015). However, these textures sit on structures that are irregular on the meter-scale and upwards, while smooth terrains are essentially flat on such size scales.

The resulting richness in shadows on the cm-dm scale in both types of terrains may explain why $q$-values from optical photometry are consistently high and do not seem to vary much from place to place on one target, or among targets. However, the heat conduction may be sufficiently effective on cm-dm size scales to smooth out the lateral temperature gradients that otherwise would form because of the shadows that the small pebbles cast (e.g., Davidsson and Rickman 2014). If so, self-heating effects (e.g., Hansen 1977) may be rather weak in smooth terrain that lacks larger-scale shadows, while it should be stronger in consolidated terrains where irregularities on the meter-scale and upward are common. Testing this hypothesis through continued analysis of VIRTIS data appears important. If it can be demonstrated that self-heating effects differs strongly between smooth and consolidated terrains, it would offer the possibility of determining the area fractions of both terrain types for a large number of unresolved comet nuclei, e.g., through observations with the James Webb Space Telescope.

## 4.2. Bulk density

### 4.2.1. State of the art before Rosetta

Before Rosetta, the density of comet nuclei could only be measured indirectly (Table 1). Solem (1995) and Asphaug and Benz (1996) modelled the tidal disruption of comet SL9 during its close approach to Jupiter, and derived densities of 550 ± 50 kg m$^{-3}$ and ~600 kg m$^{-3}$, respectively. The mass of nuclei can also be estimated based on their acceleration due to non-gravitational forces (NGF), caused by the rocket-like effect of outgassing while the comet passes through perihelion, which produce measurable changes in their orbital parameters. The density can then be calculated from this mass and the volume, the latter being estimated from resolved images for spacecraft targets or from photometry for a larger number of comets (Sosa and Fernández 2009). This method based on NGF was first used by Rickman (1986) to derive a surprisingly low density of 100 – 200 kg m$^{-3}$ for comet 1P, and after by several authors (e.g. Davidsson and Gutiérrez 2005, 2006).

Flyby missions to comets provided additional constraints on the density. The shape of the ejecta plume resulting from the Deep Impact experiment on comet 9P was used to estimate a bulk density of 470 kg m$^{-3}$, with a range of possible values between 240 and 1250 kg m$^{-3}$ (Richardson et al. 2007; Thomas et al. 2013a). Using the characteristic shape of the nucleus of comet 103P, and in particular the "waist" connecting the two lobes, A'Hearn et al. (2011) estimated a mean value for the density of 220 kg m$^{-3}$ with possible values between 140 and 520 kg m$^{-3}$ (Richardson and Bowling 2014).

Radar observations can be used to constrain the density of comets. However, it is important to be aware that the radar measurements are only sensitive to the top layer of the comets (down to the penetration depth of the radar wave; ~10 wavelengths for packed soils, i.e. up to ~10 m). The density of the surface layers can be estimated from the radar albedo if the nucleus surface is covered by a thick homogeneous layer (e.g. Harmon et al. 2004). The radar surface density estimates range between 500 and 1500 kg m$^{-3}$ (Harmon et al. 2004) and are generally larger than bulk density estimates derived from NGF, reinforces the idea that harder (denser) materials are present in the top layer (Sect. 3 and Davidsson et al. 2009).

Finally, attempts to constrain the bulk density of JFC nuclei from visible photometry (light-curves) have also been made. If the nuclei are modelled as strengthless prolate ellipsoids that are held together only by gravity, a minimum density is required to balance the centrifugal force for the given rotation rate. Considering all JFCs with available rotation rates and minimum axis ratios, Lowry and Weissman (2003) determined a cut-off in density at 600 kg m$^{-3}$, later confirmed by Snodgrass et al. (2006) and Kokotanekova et al. (2017). The lack of objects that require larger minimum densities implies that these objects have been destroyed by fast rotation (Fig. 12). In analogy with the clear spin barrier seen in asteroids (Harris 1996; Pravec et al. 2002), 600 kg m$^{-3}$ is therefore considered to correspond to the average bulk density of JFCs. While this density estimate is indirect and relies on a simplified model based on assumptions about the material strength and the shape of JFCs, the derived result is in excellent agreement with the densities estimated from other methods.

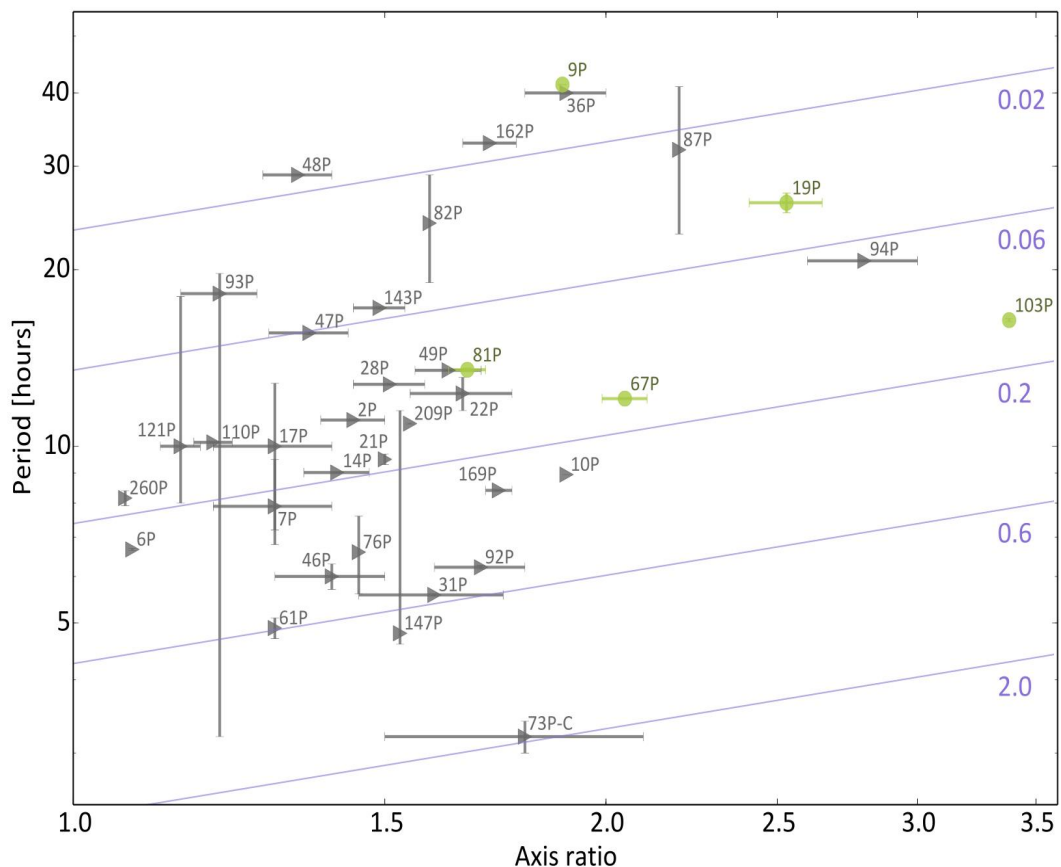

**Fig. 12** Rotation period against axis ratio for JFC nuclei (adapted from Kokotanekova et al. 2017). The triangles correspond to the lower limit of the axis ratio determined from light-curve observations. The circles denote comets visited by spacecraft. The diagonal curves show the minimum density (in g cm-3) required to keep a strengthless body with the given axis ratio and spin period stable against rotational splitting. Apart from 73P, which has recently undergone a splitting event, no other comets require densities larger than ~0.6 g cm$^{-3}$.

| Comet | Preferred value (kg m$^{-3}$) | Range (kg m$^{-3}$) | Method | Reference |
|---|---|---|---|---|
| 1P |  | 100-200 | NGF | Rickman (1986) |
|  | 600 | 200-1500 | NGF | Sagdeev et al. (1988) |
|  |  | 500-1200 | NGF | Skorov and Rickman (1999) |
|  | 1000 | 30-4900 | NGF | Peale (1989) |
|  | 500 | 200-800 | NGF | Sosa and Fernández (2009) |
| 2P | 800 | ≤1600 | NGF | Sosa and Fernández (2009) |
| 6P | 150 | 100-200 | NGF | Sosa and Fernández (2009) |
| 8P |  | 900 - 1900 | Radar | Harmon et al. (2010) |
| 9P | 400 | 200-1000 | Deep impact ejecta | Richardson et al. (2007) |
|  | 450 | 200-700 | NGF | Davidsson et al. (2007) |
|  | 200 | 100-300 | NGF | Sosa and Fernández (2009) |
|  | 470 | 240-1250 | Deep impact ejecta | Thomas et al. (2013a) |
| 10P | 700 | 300-1100 | NGF | Sosa and Fernández (2009) |
| 19P | 490 | 290-830 | NGF | Farnham and Cochran (2002) |
|  |  | 100-300 | NGF | Davidsson and Gutiérrez (2004) |
| 22P | 200 | 100-300 | NGF | Sosa and Fernández (2009) |
| 46P | 400 | 100-700 | NGF | Sosa and Fernández (2009) |
| 67P | ≤500 | 100-370 | NGF | Davidsson and Gutiérrez (2005) |
|  |  | 600-1000 | Radar | Kamoun et al. (2014) |
|  | 535 | 500-570 | In-situ mass and volume | Preusker et al. (2015) |
|  | 532 | 525-539 | In-situ mass and volume | Jorda et al. (2016) |
| 81P |  | ≤600-800 | NGF | Davidsson and Gutiérrez (2006) |
|  | 300 | ≤800 | NGF | Sosa and Fernández (2009) |
| 103P |  | 200-400 | Shape and rotation | Thomas et al. (2013b) |
|  | 220 | 140-520 | Erosion rate | Richardson and Bowling (2014) |
| D/S-L 9 | 550 | 500-600 | Tidal forces | Solem (1995) |
|  | 600 | 500-1000 | Tidal forces | Asphaug and Benz (1996) |

**Table 1** Bulk density estimates for cometary nuclei, using different methods (see text for details).

### 4.2.2. Progress during Rosetta

Rosetta provided the first direct and precise measurement of a cometary bulk density. The detailed shape model of the comet nucleus allowed its volume to be calculated with great precision (Preusker et al. 2015; Jorda et al. 2016). The mass of the comet was determined by the Radio Science Investigation (RSI) instrument on board Rosetta (Pätzold et al. 2007), where careful tracking of Rosetta's motion, via the Doppler shift in the radio link with Earth, was used to measure the gravitational influence of the nucleus on the spacecraft. Combining the two parameters allowed Jorda et al. (2016) to determine that the bulk density of the nucleus of 67P is 532 ± 7 kg m$^{-3}$. This study also determined that the nucleus has high porosity of 70 - 75%.

This method is precise enough to measure higher-order gravitational terms (i.e. the mass distribution within the comet) and the mass loss during the perihelion passage while Rosetta was in operation. These revealed an offset between the centre of figure and centre of mass, indicating subtly different densities between the "head" and "body" of 67P's nucleus (Pätzold et al. 2016).

It is worth mentioning that the bilobate nucleus of 67P would not be stable without any cohesion, and the cohesive strength has been estimated to 10 – 200 Pa at its current rotation rate of 12.4 hours by Hirabayashi et al. (2016), compatible with the strength estimates reported in Sect. 3.1.

## 4.3. Porosity from dielectric properties

The electrical parameters of a medium describe how the electric field interacts with the charges and dipoles present in the material through a number of physical processes (Kingery et al. 1976) such as electronic, atomic, dipolar and space charge polarizations, as well as the motion of free charges in matter. These processes are not instantaneous and the electrical parameters are consequently frequency dependent. At a given frequency, the electrical properties are described by the effective complex permittivity relative to vacuum, $\varepsilon = \varepsilon' - i\varepsilon''$. $\varepsilon'$ is the real part of the relative permittivity (often called the dielectric constant) and drives the velocity of the electromagnetic wave and the refraction phenomenon (in low loss media where $\varepsilon'' \ll \varepsilon'$). $\varepsilon''$ is the imaginary part of the relative permittivity and is mostly responsible for the attenuation of electromagnetic waves propagating through the medium. The complex permittivity value depends on the physical and chemical properties of the medium, as well as on its temperature.

Instruments like radars and permittivity probes make use of electromagnetic fields and aim to determine in a non-destructive way and at their operating frequency, the complex permittivity value of medium inside the sounded volume. In the case of a non-homogeneous medium, the achieved spatial resolution depends on the instrument's technical characteristics (i.e. frequency bandwidth for a radar and distance between the electrodes for a permittivity probe). Within their spatial resolution, the instruments see a homogeneous medium characterized by its bulk properties. The retrieved bulk permittivity values will then provide constraints on some properties of the material such as composition and porosity.

In the case of a comet nucleus made of ices and refractory materials (hereafter dusts), the mixture permittivity is a function of the known temperature ($T$) and frequency ($\nu$), and of the partially unknown volume fraction of ices ($x_{ices}$), dusts ($x_{dusts}$) and vacuum ($x_{vacuum}$), and their respective permittivity values ($\varepsilon_{ices}$, $\varepsilon_{dusts}$, $\varepsilon_{vacuum}$):

$$\varepsilon = f\left(\varepsilon_{vacuum}, \varepsilon_{dusts}, \varepsilon_{ices}, x_{vacuum}, x_{dusts}, x_{ices}, T, \nu\right) \quad (7)$$

where $\varepsilon_{vacuum} = 1$ and $x_{vacuum} + x_{dusts} + x_{ices} = 1$.

Different mathematical expressions for the *function* in Eq. (7) are provided by a number of so-called mixing laws, which are based on different assumptions on the density, size, shape and orientation of the inclusions (Sihvola 1999; Hashin and Shtrikman 1962; Looyenga et al. 1965; Choy 1999). This, added to the too many unknown parameters, makes the estimation of the porosity from this single equation quite impossible without additional inputs such as the dust to ice ratio, the bulk density, and the refractory materials to be considered.

An effective alternative approach to avoid the use of mixing laws is based on experimental measurements of the permittivity, performed on mixture samples in known conditions (e.g., Campbell and Ulrichs 1969; Heggy et al. 2001; Brouet et al. 2016a). But again, the same additional inputs are also needed to limit the number of samples to characterize.

In any case, knowledge of the permittivity value does provide additional constraints on the porosity value to the one obtained independently by the bulk density value. In conclusion, using both permittivity and density bulk values, and additional hypotheses about dusts to ices ratio and dusts nature, it is possible to get reliable estimates of the porosity.

**4.3.1. State of the art before Rosetta**

Earth-based radars, like the Arecibo Observatory S-band radar ($\lambda = 12.6$ cm) and the Goldstone Solar System Radar S-band ($\lambda = 12.9$ cm) and X-band ($\lambda = 3.54$ cm), are used to observe planets and small bodies, including comets from the ground. For comet nuclei, the measured radar cross-section values, together with the physical dimensions of the nucleus (that can be obtained either from radar or optical measurements), provide an estimate of the nucleus's radar albedo, which sets constraints on the dielectric properties of the sounded volume. As explained earlier, it is then possible to obtain limits on the porosity value of the nucleus subsurface. Note that the penetration depth of the radar waves depends on the radar wavelength and on the sounded matter. A rule of thumb is that, for porous media, bulk values estimated from radar data would be characteristic of the subsurface down to a depth that can reach 20 times the wavelength (i.e. 2.5 m for S-band radars).

Unfortunately, cometary nuclei are not easily detected by Earth-based radars (Ostro 1985). Their kilometric-size compared to their distance to Earth is only part of the explanation. In a number of radar observations, the echo from cm-size grains in the coma prevented the detection of the weak nucleus echo. Nevertheless, in most cases, even the non-detection of the nucleus by radar still allowed limits of the nucleus radar albedo to be set, and to constrain the density or porosity of the nucleus top layer.

The first successful radar detection of a comet nucleus was made in 1980 of comet 2P/Encke with the Arecibo S-band radar (Kamoun et al. 1982), followed by a series of observations leading to the detection of more than 14 nuclei and the non-detection of several of them (e.g., Campbell et al. 1989, Harmon et al. 1997, Harmon and Nolan 2005, Harmon et al. 2011).

As explained in Sect. 4.2, most of the density values estimated from nucleus radar echoes lie between 0.3 and 0.7 g cm$^{-3}$ (Table 1), which is roughly consistent with a high porosity of the subsurface. Comet 8P/Tuttle stands out though, with an estimated density at the subsurface significantly higher at between 0.9 and 1.9 g cm$^{-3}$ (Harmon et al. 2010).

The case of 67P is of upmost interest because its nucleus has been thoroughly analysed later during the Rosetta mission. The 2.380 GHz Arecibo radar has been operated in 1982 to observe 67P at a geocentric distance of 0.4 a.u, but no detection of the nucleus or the coma was achieved then (Kamoun 1998). The lack of radar detection provided an upper limit of the nucleus radius of around 3 km (Kamoun 1983; Kamoun et al. 1998) and an upper value of the nucleus albedo at 0.05. Kamoun et al. (2014) revisited the same radar data set to provide some constraints on the nucleus properties that would be useful for the Philae landing to come. Considering that, from measurements performed on other cometary nuclei, the lower limit of radar albedo values is likely to be 0.04, this led, for 67P, to an estimated permittivity value between 1.9 and 2.1, converted to a 0.6 to 1 g cm$^{-3}$ density at the surface, compatible with 55 – 65 % porosity.

**4.3.2. Progress during Rosetta**

Two experiments of the Rosetta mission were able to sound beyond the nucleus surface and to provide estimates for the dielectric constant of the nucleus using electromagnetic waves: the Permittivity Probe of the Surface Electric Sounding and Acoustic Monitoring Experiment (SESAME-PP; Seidensticker et al. 2007) and the Comet Nucleus Sounding Experiment by Radiowave Transmission (CONSERT; Kofman et al. 1998, 2007).

SESAME-PP consisted of five electrodes operating between 10 Hz and 10 KHz, with three transmitting electrodes (located on the +X-foot, MUPUS-PEN and APXS), able to inject a current in the material in contact with the electrodes, and two receiving electrodes (located on the +Y-foot and -Y-foot). However, data processing revealed that only the range of frequency between 409 and 804 Hz was usable to derive a constraint on the dielectric constant of the material, and that only one transmitter could be used (+X-foot). The measurements performed by SESAME-PP were restricted approximately to the first meter of cometary material below the final Abydos landing site of Philae (Lethuillier et al. 2016).

Lethuillier et al. (2016) determined a lower limit of 2.45 ± 0.20 for the dielectric constant.

CONSERT was a bistatic radar instrument. The principle of measurement was based on the transmission and reception of a radio signal between the orbiting Rosetta spacecraft and the Philae lander. It operated at a centre-band frequency of 90 MHz, so that it propagated long-wavelength electromagnetic waves through part of the cometary nucleus. In a low loss medium, the measurement of the signal travel time allows the determination of the wave velocity and thus to estimate the real part of the permittivity value. CONSERT was able to provide a dielectric constant equal to 1.27 ± 0.05 from signals that propagated through the small lobe of the nucleus over distances of hundreds of meters (Kofman et al. 2015).

The very low value of CONSERT contrasts with the lower limit determined by SESAME-PP, although both results indicate a dielectric constant representative of a highly porous material (i.e., a volume fraction of voids larger than 50%; e.g., Heggy et al. 2012; Brouet et al. 2014, 2015, 2016a). The temperature of the near-surface material was low enough during the measurement sequences, i.e. lower than ~180 K, to assume a nondispersive behaviour of the dielectric constant (Lethuillier et al. 2016) in the operating frequency range of SESAME-PP and CONSERT, allowing the discrepancy between the near-surface dielectric constant and the internal dielectric constant to be only caused by a variation of the porosity and/or the composition. Before the final results released by the SESAME-PP science team regarding the measurement of the dielectric constant, Ciarletti et al. (2015) had already pointed out that the dielectric constant may vary as a function of depth within the shallow subsurface of the Abydos landing site, suggesting a decrease of its value with depth, and excluding a significant increase of the bulk density or of the dust-to-ice mass ratio with depth in the first meters.

In order to interpret the dielectric constant value in terms of porosity, two independent methods were used:

1) A first method is based on the use of dielectric mixing laws previously mentioned. Kofman et al. (2015) used a conservative approach by using the Hashin-Shtrikman bounds that encompass all the solutions suggested by the numerous existing dielectric mixing laws. Kofman et al. (2015) also compiled laboratory dielectric data on relevant refractory materials and ices from the existing literature in order to derive the porosity, and finally, found a porosity of 75% – 85% with this method. Lethuillier et al. (2016), following the same method, suggested that a lower limit of 2.45 ± 0.20 for the dielectric constant is compatible with a porosity lower than 50% if carbonaceous chondrites are taken into account as the dust component of the ternary mixture (we note that ordinary chondrites have been excluded as potential dust analogues to explain the dielectric constant estimated by CONSERT, see Kofman et al. 2015). With the same method of analysis, Hérique et al. (2016) refined the possible range of the zero-porosity dielectric constant of the dusty component of 67P/CG.

2) A second method is based strictly on laboratory dielectric measurements of cometary analogues performed in a frequency range encompassing the CONSERT operating frequency, i.e. between 50 MHz and 2 GHz (Brouet et al. 2015, 2016a). The studied samples included pure laboratory-grown water ice and JSC-1A Lunar simulants, as well as mixtures of these two components, with controlled porosities in the range of 30% to 91% and dust-to-ice mass ratios in the range of 0 to 1. From these measurements, a semi-empirical dielectric law has been derived and used to estimate a porosity lower than 58%, and between 75% and 86%, from the values of the dielectric constant measured by SESAME-PP and CONSERT, respectively (Brouet et al. 2016b). Thus, the results suggest an increase of the porosity with depth, like the first method, and also indicate that this porosity gradient over the small lobe of the nucleus cannot be excluded even when taking into account the variation of the dust-to-ice mass ratio.

Penetrating radars like CONSERT are able to detect structures of spatial scales much larger than the wavelength (i.e. several tens of meters), which do not seem to be present in the small lobe sounded by the CONSERT radar. At a much smaller scale, commensurate with the wavelength, the analysis of the shape of the received pulses provides information about the potential heterogeneities inside the sounded volume. Since no significant modification of the pulse width has been observed (Ciarletti et al. 2017), it is possible to conclude that no volume scattering took place. Based on extensive simulations of propagation through non-homogeneous media with different spatial characteristic scales and permittivity values, the presence of strong heterogeneities at a scale of a few meters has been ruled out.

## 4.4. Synthesis on structure and vertical stratigraphy

Rosetta investigated the interior of 67P's nucleus on scales ranging from millimetres to kilometres, using electromagnetic waves from the kilohertz via the gigahertz and infrared to the visible light range, its gravity field, and elastic waves. Some instruments provided mappings of large areas, others focused on isolated spots. Putting all parts together into a single stratigraphy requires some simplifications, but nevertheless all of the investigations reported here fit together into a coherent picture (Fig. 13).

The broad picture is that the bulk of the comet consists of a weakly bonded, rather homogeneous material that preserved primordial properties under a thin shell of processed material, and possibly covered by a granular material that is able to form pseudo-aeolian morphologies like dunes and moats (El-Maarry et al. 2018). This cover might in places reach thicknesses of many meters.

The orientations of subhorizontal and subvertical surfaces indicates the presence of an onion-like layering in both lobes of 67P, but like layering in snow or sandstone, the associated changes of physical properties are smaller than the resolution of the methods available with Rosetta. There are no indications of strong nucleus heterogeneities at a scale of a few meters on 67P's small lobe.

The processed outer shell is the result of competing effects, resulting from the sublimation of volatiles and super volatiles heated up for the first time since their aggregation, which on the one hand remove and eject material into space, but on the other hand re-condensate between grains and thus cement the shell.

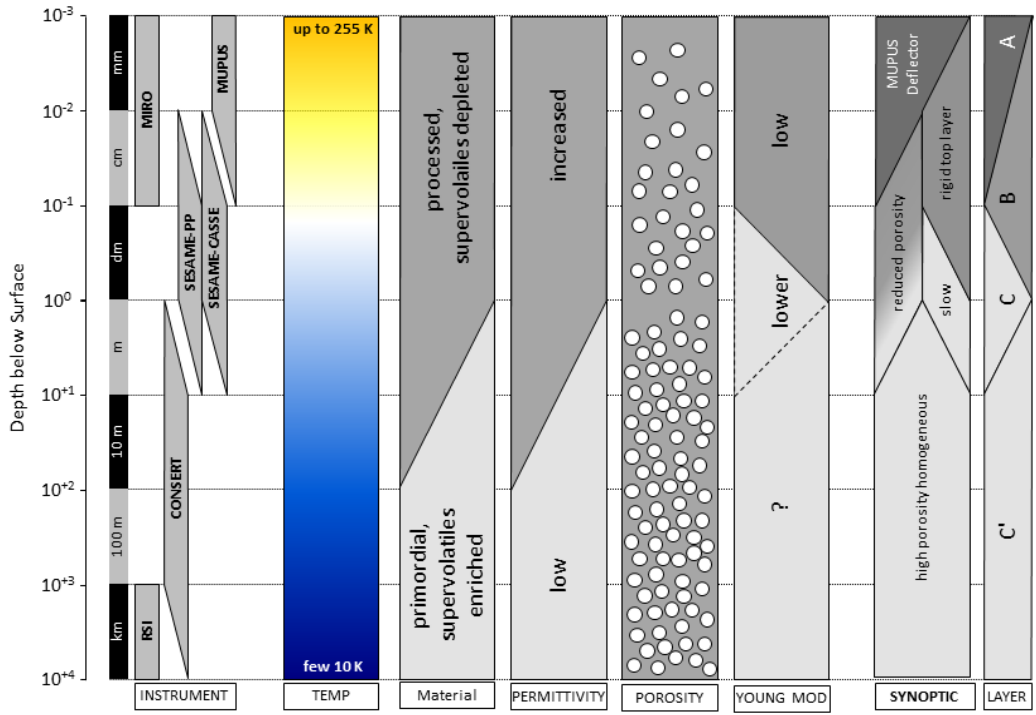

**Fig. 13** Synthesis of the (simplified) stratigraphy of 67P/Churyumov-Gerasimenko on a logarithmic depth scale (granular cover found on subhorizontal surfaces not shown). Columns from left to right: (1) instruments providing observations of the subsurface, (2) temperature range, indicating near-perihelion surface temperatures and estimated temperature of the deep interior, (3) materials, simplified into primordial and not primordial, (4), electrical permittivity resulting from radar and in-situ measurements, (5) porosity, resulting from electrical properties and bulk density, (6) Young's modulus from SESAME/CASSE, (7) and (8) synoptic view – Layer A: MUPUS deflector is a possible thin strong layer that prevented MUPUS from penetrating, but is too thin to be seen with CASSE; Layer B: Rigid top layer; Layer C "slow" (=low shear wave velocity) layer derived from CASSE, porosity gradient in both layers B and C; Layer C' is the deep primordial material, layers C and C' might be identical.

## 5. Conclusions

We have presented a review of the thermal, mechanical, structural, and dielectric properties of cometary nuclei, after Rosetta, including more than 200 references. Our work includes several syntheses on the thermal properties (Sect. 2.3), mechanical properties (Sect 3.3) and structural and dielectric properties (Sect. 4.4), with key figures (Figs. 6, 10 and 13). Here are the main conclusions for cometary nuclei:

**Temperature** – The temperature varies across the surface, following solar insulation to first order. On the dayside, there exist some deviations from the black body temperature due to surface roughness, self-heating, and heat conductivity. On the night side, the temperature is mainly controlled by heat conductivity. At the ~10 m px$^{-1}$ spatial resolution, it is not yet demonstrated that the presence of water ice reduces significantly the surface temperature. Because of the low thermal inertia, the nucleus temperature decreases very rapidly with depth, from hundreds of Kelvin to tens of Kelvin in the first meter.

**Thermal inertia** (Fig. 6) – The thermal inertia is in the range 0 – 350 J K$^{-1}$ m$^{-2}$ s$^{-1/2}$, with a best estimate of 0 – 60 J K$^{-1}$ m$^{-2}$ s$^{-1/2}$ for the bulk value. On 67P, most values are in the range 10 – 170 J K$^{-1}$ m$^{-2}$ s$^{-1/2}$, with variations across the surface: the thermal inertia of smooth terrains covered by deposits is lower (typically <30 J K$^{-1}$ m$^{-2}$ s$^{-1/2}$) than that of exposed consolidated terrains (typically >110 J K$^{-1}$ m$^{-2}$ s$^{-1/2}$) (Leyrat et al. 2015).

**Thermal conductivity** – We derive a thermal conductivity of 0 – 0.009 W m$^{-1}$ K$^{-1}$, from the above thermal inertia range of 0 – 60 J K$^{-1}$ m$^{-2}$ s$^{-1/2}$, assuming a nucleus with a density of 532 kg m$^{-3}$ (Jorda et al. 2016) and a heat capacity of 770 J kg$^{-1}$ K$^{-1}$ (dust at 300 K; Winter and Saari 1969).

**Tensile and compressive strength** (Fig. 10) – The global tensile strength is <100 Pa. This low value is consistent with laboratory experiments and modelling, when scaled to the meter scale. On 67P, the strength varies across the surface, and there exists locally a sub-surface layer of "harder" materials with a compressive strength of 3.5 – 12 kPa at Agilkia (first Philae touchdown) and 18 – 9800 kPa at Abydos (final Philae landing). At Abydos, this layer is shallow (10's of cm) and compares well with porous (75%) snow or silica aerogel, in terms of mechanical properties.

**Elastic properties** – The Young's modulus is only known for 67P at the Abydos final landing site and amounts to 7.2 – 980 Mpa (Knapmeyer et al. 2018). This value applies to the top layer, up to a depth of 0.1 m to 0.5 m, below which the Young's modulus is reduced, most likely by the presence of softer, more porous, materials.

**Surface roughness** – At the >10 m scale, their exist a large variety of terrains with different degrees of roughness (smooth or irregular), observable visually on spatially resolved images. At the <10 m scale, i.e. spatially unresolved excepted for 67P, small-scale roughness yields to Hapke mean slope angles $\theta$ in the range 5 – 32°, with variations from one comet to the other. On 67P, close-up images reveal that smooth terrains are mostly pebbles (cm-dm scale), while consolidated terrains consist of cemented/sintered agglomerates of mm-cm sized grains and fine-grained material (Bibring et al. 2015).

**Bulk density** – The mean value for the bulk density is 480 ± 220 kg m$^{-3}$, based on 20 estimates (Table 1), with a range between 150 kg m$^{-3}$ and 1000 kg m$^{-3}$. The most accurate estimate is for 67P, with 532 ± 7 kg m$^{-3}$ (Jorda et al. 2016).

**Dielectric properties** – The permittivity of the nucleus subsurface layer, derived from radar albedo, is most likely in the range 1.9 – 2.1 (Kamoun et al. 2014). For 67P, at the final landing site, the dielectric constant was estimated to a lower limit of 2.45 ± 0.20 by Lethuillier et al. (2016) with SESAME-PP and to 1.27 ± 0.05 by Kofman et al. (2015) with CONSERT. This discrepancy suggests a decrease of the dielectric constant with depth, excluding a significant increase of the bulk density or of the dust-to-ice mass ratio with depth in the first meters (Ciarletti et al. 2015).

**Porosity** – The low density of cometary nuclei suggests a high porosity (70 – 80 %) for most of them. For 67P, adding the constraint from the dielectric properties, the porosity increases from 58% in the first meter to 75 – 86% below.

**Vertical stratigraphy** (Fig .13) – The broad picture is that the bulk of the comet consists of a weakly bonded, rather homogeneous material that preserved primordial properties under a thin shell of processed material, and possibly covered by a granular material. This cover might in places reach thicknesses of many meters. The processed outer shell is the result of competing effects between (i) the sublimation of volatiles that ejects material into space, and (ii) the recondensation of volatiles between grains that cement the shell. The physical properties of the top layer (the first meter) are not representative of that of the bulk nucleus; the top layer is always thermally processed, and moreover denser, less porous, and harder when made of consolidated materials. Finally, strong nucleus heterogeneities at a scale of a few meters are ruled out in 67P's small lobe.

All these physical properties put constraints on the formation and evolution scenarios of cometary nuclei. Comets are among the least dense bodies of the solar system, highly porous and very weak, which have been able to preserve highly volatiles species (e.g. CO, $CO_2$, $CH_4$, $N_2$) into their interior since their formation, protected from solar heating by a low bulk thermal conductivity. This points toward a formation and evolution scenario that avoids global heating above ~90 K (amorphous/crystalline water ice transition) and global mechanical alteration of the nucleus. As discussed in Weissman et al. (2018), a formation by hierarchical agglomeration (Davidsson et al. 2016) or a formation by pebbles accretion in streaming instabilities (Blum et al. 2014) are both possible. After its formation, 67P either remained intact until today, in which case it would be a primordial object (Davidsson et al. 2016), or it experienced one or several sub-catastrophic collisions (Jutzi et al. 2017) or catastrophic disruptions (Schwartz et al. 2018) with the re-aggregation of materials as rubble-pile. Alternatively, 67P could also be the fragment of a larger Kuiper belt object (Davis and Farinella 1997; Morbidelli and Rickman 2015), but this hypothesis is likely limited to a parent body not exceeding 10 km in size. Indeed, as explained by Davidsson et al. (2016), only small bodies (diameter <10 km) formed by the slow (~3 Myr) hierarchical agglomeration of materials can preserve a temperature <100 K in their interior, even in the presence of radioactive heating, allowing them to keep their super volatiles, most probably trapped in amorphous ice or clathrates (Mousis et al. 2015). Larger bodies (50 – 100 km in size) have a lithostatic pressure of several to tens of kPa, much larger than the inferred tensile strength of <100 Pa, which would mean that, in this collisional fragment scenario, 67P would be made of numerous mechanically strong (kPa – MPa) blocks and boulders, assembled together for a resulting low bulk tensile strength (<100 Pa);

additionally, preserving the highly volatiles species during the giant catastrophic collision (~0.5 km s$^{-1}$) of two large bodies appears challenging. Disentangling between the different above scenarios however goes beyond the scope of this paper and we refer to Weissman et al. (2018) for a more detailed discussion of the pros and cons of each of them.

There is no doubt that space missions and Rosetta in particular have greatly improved our knowledge on comets, thanks to the missions themselves but also to the numerous parallel studies they motivated. For future space missions, the major challenge concerning the nucleus physical properties is probably to interact with the surface in a more controlled way than Philae did, and to accurately probe the deep interior. For ground and space telescopes, especially important due to the scarcity of in-situ space missions, the challenge will be to use the knowledge gained recently to better characterize the thermal, mechanical and structural properties of a large number of cometary nuclei, even possibly resolving spatially the largest and closest ones with forthcoming thirty-meter class telescopes (providing a spatial resolution <1 km at 0.3 au from the Earth).

## Acknowledgments

This project has received funding from the European Union's Horizon 2020 research and innovation programme under grant agreement no. 686709. This work was supported by the Swiss State Secretariat for Education, Research and Innovation (SERI) under contract number 16.0008-2. The opinions expressed and arguments employed herein do not necessarily reflect the official view of the Swiss Government. Yuri Skorov thanks the Deutsche Forschungsgemeinschaft (DFG) for support (grant SK 264/2).